\documentclass[11pt]{article}
\usepackage{jheppub}
\usepackage{amsmath,amssymb,amsfonts,graphicx,slashed,amsthm,mathtools,upgreek, enumerate, tensor, subfig}
\usepackage[dvipsnames]{xcolor}
\usepackage{arydshln}
\usepackage{slashed}

\usepackage{comment}
\usepackage{hyperref}
\usepackage[utf8]{inputenc}
\usepackage[titletoc]{appendix}

\numberwithin{equation}{section} 
\allowdisplaybreaks

\newcommand{\beq}{\begin{equation}}
\newcommand{\eeq}{\end{equation}}
\newcommand{\bea}{\begin{equation}\begin{aligned}}
\newcommand{\eea}{\end{aligned}\end{equation}}
\newcommand{\bra}[1]{\langle #1 |}
\newcommand{\ket}[1]{| #1 \rangle}

\newcommand{\avg}[1]{\langle #1 \rangle}

\DeclareMathOperator{\tr}{tr}

\title{Spin structures and baby universes}

\author[a,b]{Vijay Balasubramanian}
\author[a]{\!, Arjun Kar}
\author[c]{\!, Simon F. Ross}
\author[d,e,a]{\!, Tomonori Ugajin}

\affiliation[\,a]{David Rittenhouse Laboratory, University of Pennsylvania,\\
209 S.33rd Street, Philadelphia, PA 19104, USA}
\affiliation[\,b]{Theoretische Natuurkunde, Vrije Universiteit Brussel (VUB), and \\ International Solvay Institutes, Pleinlaan 2, B-1050 Brussels, Belgium}
\affiliation[\,c]{Centre for Particle Theory, Department of Mathematical Sciences, Durham University,\\
South Road, Durham DH1 3LE, UK}
\affiliation[\,d]{Center for Gravitational Physics,
Yukawa Institute for Theoretical Physics, Kyoto University,\\
Kitashirakawa Oiwakecho, Sakyo-ku, Kyoto 606-8502, Japan}
\affiliation[\,e]{The Hakubi Center for Advanced Research, Kyoto University,
Yoshida Ushinomiyacho, Sakyo-ku, Kyoto 606-8501, Japan}

\emailAdd{vijay@physics.upenn.edu}
\emailAdd{arjunkar@sas.upenn.edu}
\emailAdd{s.f.ross@durham.ac.uk}
\emailAdd{tomonori.ugajin@yukawa.kyoto-u.ac.jp}

\abstract{We extend a  2d topological model of the gravitational path integral to include sums over spin structure, corresponding to Neveu-Schwarz (NS) or Ramond (R) boundary conditions for fermions.  The Euclidean path integral  vanishes when the number of R boundaries is odd.  This path integral corresponds to a correlator of boundary creation operators on a non-trivial baby universe Hilbert space. The non-factorization necessitates a dual interpretation of the bulk path integral in terms  of a product of partition functions (associated to NS boundaries) and Witten indices (associated to R boundaries), averaged over an ensemble of  theories with varying Hilbert space dimension and different numbers of bosonic and fermionic states.  We also consider a model with End-of-the-World (EOW) branes: the dual ensemble then includes a sum over randomly chosen fermionic and bosonic states.  We propose two modifications of the bulk path integral which  restore an interpretation in a single dual theory: (i) a geometric prescription where we add extra boundaries with a sum over their spin structures, and (ii) an algebraic prescription involving ``spacetime D-branes". We extend our ideas to Jackiw-Teitelboim gravity, and propose a dual description of a single unitary theory with spin structure in a system with eigenbranes. 
}


\begin{document}

\maketitle

\parskip=10pt

\section{Introduction}

\begin{figure}[t]
    \centering
    \includegraphics[scale=.3]{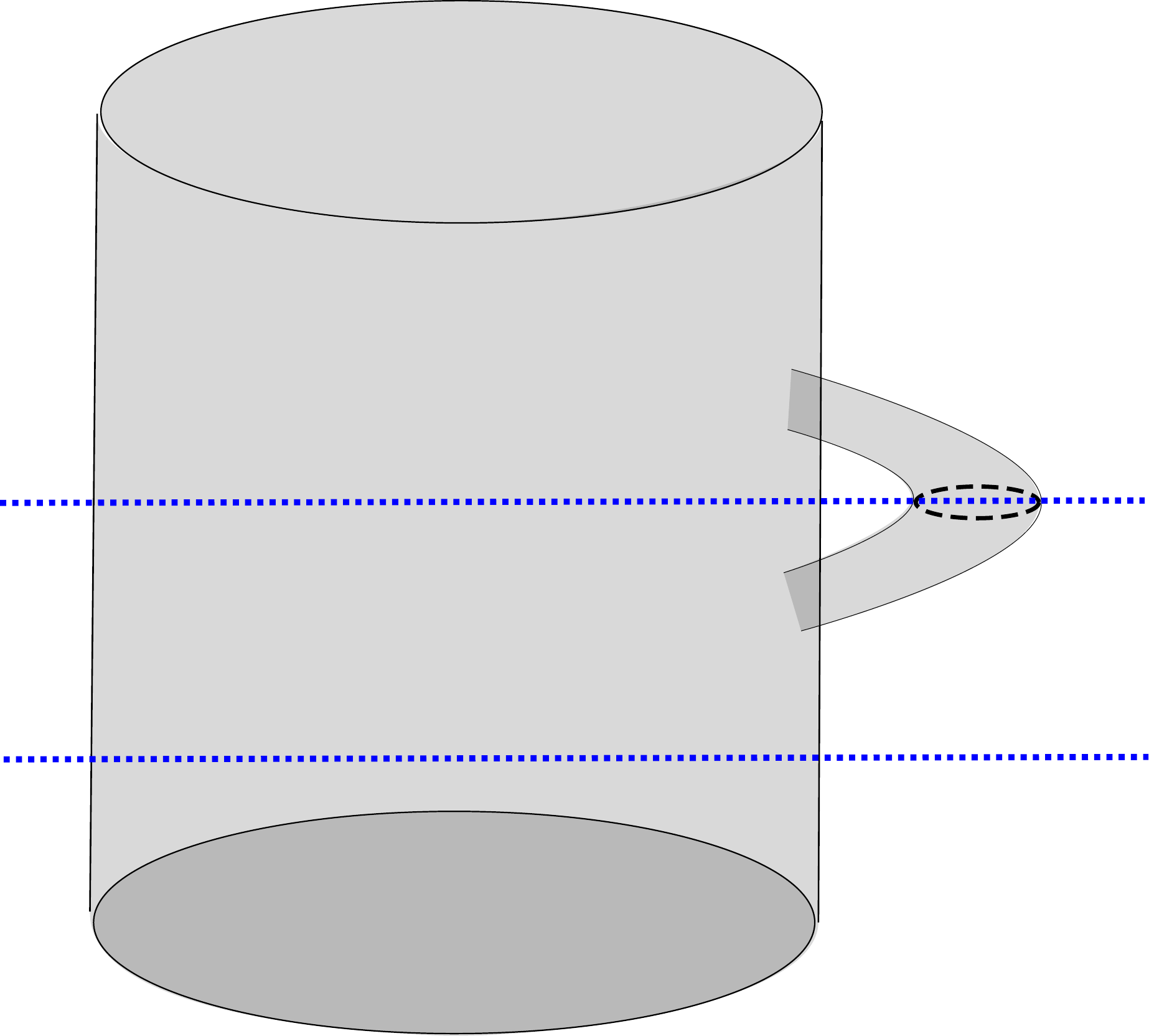}
    \caption{\small{A Euclidean spacetime with a ``wormhole" handle which contributes to the gravitational path integral.   If we slice the path integral open along the bottom blue dotted line, we get a state in a universe with a connected Cauchy slice.  If we instead slice along the top blue dotted line, we see the parent universe has ``emitted" a baby universe, which appears as a disconnected component of the Cauchy slice.}}
    \label{fig:baby-universe}
\end{figure}

The Euclidean gravity path integral naturally includes a sum over topologies of the spacetime manifold. A semiclassical saddlepoint contribution then has, in addition to the fluctuations of the geometry and fields on the given background topology, additional fluctuations associated with changes in the topology. A simple kind of topology changing process is the addition of a small handle to the geometry (figure \ref{fig:baby-universe}). 
Such a handle is referred to as a spacetime wormhole, and is distinct from the more familiar spatial wormhole where the spatial geometry at any moment in time has a handle. If we slice across the picture through the handle, we see that it can be interpreted as the emission of a small ``baby universe" by a parent spacetime. 

The role played by these topology changing processes in unitary quantum gravity is an old puzzle \cite{Lavrelashvili:1987jg,Hawking:1987mz,Giddings:1987cg,Hawking:1988ae}. In the 1980's, Coleman \cite{Coleman:1988cy} and Giddings and Strominger \cite{Giddings:1988cx,Giddings:1988wv} proposed that at scales large compared to the size of the baby universes, the effects on the parent universe are encoded in an average over a set of couplings, referred to as alpha parameters, for local operators in the parent universe.

A different issue with spacetime wormholes appears in the AdS/CFT correspondence, where the partition sum of the boundary field theory is computed by the bulk Euclidean gravitational path integral.  In this context, we expect that independent, non-interacting field theories in their vacuum state  will be dual to a sum over disconnected bulk geometries, because the correlations  functions should factorize between the theories. However, it turns out that there are examples of spacetimes with two asymptotic boundaries for which there are solutions where the two boundaries are connected by a spacetime wormhole, in addition to the disconnected geometries connected to the two boundaries separately \cite{Maldacena:2004rf}. The usual rules of the AdS/CFT correspondence state that we should sum over {\it all} bulk manifolds that fill in the boundaries on which the field theories are defined, so we expect to add the contribution of the wormhole to that of the disconnected geometries which are asymptotic to the two spacetime boundaries.\footnote{This is different from the case of the Hawking-Page transition in AdS/CFT where two Euclidean bulk geometries, thermal AdS and the Euclidean black hole, compete to dominate the partition sum calculation for the dual thermal field theory.  In this case, in the black hole phase a Cauchy slice  has a wormhole between the two asymptotic regions of the eternal black hole, but the  boundary of the Euclidean saddlepoint has only one connected component.  By contrast, in the spacetime wormholes, the Euclidean geometry has multiple disconnected boundary components.} The contributions of the wormholes in the gravitational path integral suggests a connection between the field theories on the two boundaries, and a lack of factorization of their correlation functions.  This is puzzling because we supposedly started with two independent field theories in their vacuum state whose correlation functions should have factorized.  This example suggests a subtlety in the quantity that is actually computed by the gravitational path integral in the AdS/CFT correspondence and raises the question of whether it is simply the vacuum partition function.\footnote{In principle, we could define the gravitational path integral to only include disconnected bulk manifolds, and in this way enforce that the bulk path integral should factorize.  However, recent work suggests that including the non-factorizing wormhole contributions leads to a formalism which is powerful enough to resolve the black hole information paradox \cite{Hawking:1974sw} using only semiclassical gravity \cite{Almheiri:2019qdq,Almheiri:2019hni,Penington:2019kki} in any dimension \cite{Almheiri:2019psy,Balasubramanian:2020hfs}.  See \cite{Almheiri:2020cfm} for a review.  Also, see \cite{Polchinski:1994zs} for an early application of baby universes to the information paradox. Spacetime wormholes have also led to attempts at solving the cosmological constant problem \cite{Coleman:1988tj,Preskill:1988na,Klebanov:1988eh}. As such, it is clearly worth understanding the detailed implications of these contributions to the gravitational path integral.
} 

This tension was recently resolved in the context of two-dimensional Jackiw-Teitelboim (JT) gravity in \cite{Saad:2019lba}. The authors of \cite{Saad:2019lba} showed that the gravitational path integral for JT gravity was dual not to a single boundary theory, but to a matrix model which could be interpreted as an ensemble average over a family of Hamiltonians for a one-dimensional quantum mechanics on the boundary.\footnote{Attempts to build a 2d boundary ensemble model for 3d pure gravity have been made recently in \cite{Maloney:2020nni,Afkhami-Jeddi:2020ezh,Belin:2020hea,Cotler:2020ugk}, and the existence of the ensemble in any dimension was interpreted as a resolution of the so-called ``state paradox" in \cite{Bousso:2020kmy}.} The connected correlation function of  boundary observables in this ensemble average is generated in the bulk path integral by a contribution from connected geometries with multiple boundaries, i.e., geometries with spacetime wormholes. This observation raised the question of whether there is a general connection between spacetime wormholes in the Euclidean gravity path integral and a dual holographic description in terms of ensemble-averaged quantum theories.

In \cite{Marolf:2020xie}, Marolf and Maxfield used a baby universe picture to propose that the gravitational path integral should generically have an interpretation in terms of an ensemble average. The general discussion was illustrated by considering a simple exactly solvable topological model where the bulk path integral sums over all smooth two-dimensional manifolds with $n$ boundaries, including both connected and disconnected manifolds, with a weight depending only on the topology of the bulk manifold. The connected contributions imply that the bulk path integral does not factorize: the $n$-boundary result is not a power of the one-boundary result. 

In AdS/CFT we would expect this bulk path integral to be dual to some quantum mechanical system living on the boundary. Each boundary is a Euclidean circle, so the path integral in the dual quantum theory on this circle computes the partition function $Z$.\footnote{As there is no metric in this model, there is no notion of the length of the circle, and hence no temperature. The partition function is just a number.} Marolf and Maxfield explicitly calculated the bulk path integral in this model, and showed that the path integral with $n$ boundaries can be interpreted as an ensemble average $\langle Z^n \rangle$, where $Z = \text{tr}_{\mathcal H}(1)$ is the trace over all states in a boundary Hilbert space $\mathcal H$ of dimension $d$ with vanishing Hamiltonian. Each boundary is associated with an independent copy of $\mathcal H$, and the average is taken in an ensemble of Hilbert spaces with a Poisson distribution over $d$. In the context of the ensemble of boundary theories, $Z$ should not be regarded as a number in a single theory, but rather as a random variable taking values in the set of boundary Hilbert space dimensions. As such, because of the average over the dimension $d$, $\langle Z^n \rangle \neq \langle Z \rangle^n$. 

This non-factorization can also be described in terms of a baby universe Hilbert space. We define this by slicing open the bulk path integral, along some one-dimensional surface which splits the bulk manifold into two pieces.\footnote{Since the bulk manifold can have multiple connected components, so will this surface, and its topology will differ in different contributions to the path integral.} We think of the path integral on one side of this surface as defining a state on the surface, and the full path integral as defining an inner product between two such states. The path integral on one side of the slice is completely characterised by the number $m$ of asymptotic boundaries on this side, so we denote the resulting state as $|Z^m \rangle$. If we take a path integral with $m+n$ boundaries and slice it so $m$ boundaries are on one side and $n$ on the other, this defines the overlap $\langle Z^n | Z^m \rangle$.  If we change the slicing so that $m+1$ of the boundaries now lie on one side of the slice, this defines a state $|Z^{m+1} \rangle$ and the full path integral calculates $\langle Z^{n-1} | Z^{m+1} \rangle$. It is therefore natural to define a Hermitian operator $\hat Z$  on the baby universe Hilbert space such that $\hat Z |Z^m \rangle= |Z^{m+1} \rangle$, describing the effect on the state of adding a boundary in the path integral on one side of the slice. Then the bulk path integral with $n$ boundaries can be described as an expectation value in the baby universe Hilbert space, $\langle \text{HH} | \hat Z^n | \text{HH} \rangle$, where $| \text{HH} \rangle  = | Z^0 \rangle$ is the Hartle-Hawking state, defined by the path integral with no asymptotic boundaries \cite{Hartle:1983ai}.   This correlator would factorize, i.e., $\langle \text{HH} | \hat Z^n | \text{HH} \rangle = \langle \text{HH} | \hat Z | \text{HH} \rangle^n$, if $|\text{HH}\rangle$ were an eigenstate of $\hat Z$, which is only possible if the baby universe Hilbert space $\mathcal H_{BU}$ is one-dimensional. Thus, \cite{Marolf:2020xie} showed that the bulk path integral requires a dual  interpretation in an ensemble of theories (as opposed to a single theory) precisely when the baby universe Hilbert space is non-trivial.\footnote{As pointed out in \cite{Anous:2020lka}, a key feature of the Hartle-Hawking density matrix which seems to lead to a non-trivial baby universe Hilbert space is the lack of ``bra-ket" wormholes, which are wormholes between the spacetime boundaries associated with the bra and ket in a density matrix like $\ket{Z}\bra{Z}$.}

In general then, in the simple topological model of \cite{Marolf:2020xie}, the gravitational path integral with $n$ boundaries computes a non-factorizing correlator of $n$ baby universe creation operators $\hat{Z}^n$, and this quantity has a dual description as the average of the $n^{\text{th}}$ power of the partition sum evaluated in a particular ensemble of theories: $\langle \text{HH} | \hat Z^n | \text{HH} \rangle = \langle Z^n \rangle$, where the average on the right hand side is taken over a statistical ensemble of field theories, and $Z^n= \text{tr}_{\mathcal H}(1) \times \cdots \times \text{tr}_{\mathcal H}(1)$  is the partition function in the $n$-fold product of identical theories.

A natural extension of this  model is to add spin structures, the topological aspect of fermion fields. The main aim of the present paper is to explore this extension. The addition of spin structures for JT gravity was considered in \cite{Stanford:2019vob} and our discussions have several parallels with that work.

In section \ref{top} we briefly review the topological model of \cite{Marolf:2020xie}, and introduce an extension to include a sum over spin structures. (In this section we consider a sum which weights all spin structures equally.) This extends the model to include two types of boundaries, corresponding to anti-periodic or periodic boundary conditions for fermions on the boundary circle, which are respectively dual to the Neveu-Schwarz (NS) or Ramond (R) sectors of the boundary quantum theory. If we have an odd number of Ramond boundaries, the bulk path integral vanishes identically, as there are no bulk spin structures compatible with this boundary condition. As before, we can regard the path integral, sliced open so that it has $m$ NS and $\tilde{m}$ R boundaries, as defining a state $|Z^m \tilde{Z}^{\tilde{m}}\rangle$.   Likewise, the full path integral with $n=m + m'$ and $\tilde{n}=\tilde{m}+ \tilde{m}'$ boundaries of each kind defines the overlap $\langle Z^{m'} \tilde{Z}^{\tilde{m}'}|Z^{m} \tilde{Z}^{\tilde{m}}\rangle$.  In terms of operators $\hat{Z}$ and $\hat{\tilde{Z}}$ that act on the baby universe Hilbert space to add NS and R boundaries, the full path integral computes the  correlator $\langle \text{HH}| \hat{Z}^{n} \hat{\tilde{Z}}^{\tilde{n}}|\text{HH} \rangle$ where $\ket{\text{HH}}$ is the Hartle-Hawking wavefunction.  The vanishing of these correlators for odd, but not even, $\tilde{n}$  is a particularly simple manifestation of the failure of factorization, here due to a topological obstruction, and suggests that the dual field theory defined on $n + \tilde{n}$ disconnected boundary circles must be given an ensemble interpretation.

To construct this ensemble interpretation, we consider a quantum theory with a finite-dimensional Hilbert space $\mathcal{H}$ and vanishing Hamiltonian defined on each boundary. The path integral on a circle with NS boundary conditions computes the partition function $Z = \mathrm{tr}_{\mathcal H} (1)$ while, with R boundary conditions, it computes the Witten index $\tilde Z = \mathrm{tr}_{\mathcal H} ((-1)^F)$.  We show that the gravitational path integral with $n$ and $\tilde{n}$ boundaries of each kind computes the ensemble average $\langle Z^n \tilde{Z}^{\tilde{n}}\rangle$ in the boundary dual theory, where the ensemble sums over the dimension of the boundary Hilbert spaces with Poisson weighting, and over the fraction of bosonic and fermionic states in each theory with binomial weighting. 
Note that while $\tilde Z$ vanishes on average (after summing over the fraction of fermionic states), it does not vanish in each theory individually.

As in \cite{Marolf:2020xie}, to construct the Hilbert space of baby universes, we act with formal polynomials of $\hat{Z}$ and $\hat{\tilde{Z}}$ on $\ket{\text{HH}}$. This generates a substantial number of null states, which we must quotient out. Polynomials in the Ramond boundary creation operator $\hat{\tilde{Z}}$ give an especially simple illustration of how and why these null states appear.  Null states of this kind usually signal a gauge redundancy that sharply reduces the number of physical states. We discuss the boundary interpretation of null states associated to the $\hat{\tilde{Z}}$ operator in terms of the vanishing of conditional expectation values in the dual ensemble.

In section \ref{eow}, we define a gravitational path integral in which we include end of the world (EOW) branes in the model as a simple form of topological ``matter''. These have been used to represent gravitational microstates in approaches to the black hole information paradox \cite{Penington:2019kki}, by introducing a  flavor index $i$ on the EOW branes. EOW branes can intersect the spacetime boundary; the corresponding state in the boundary theory is denoted $\psi_i$. The boundary conditions for the topological model then include circular boundaries and boundary segments terminated by EOW branes at both ends. In the model of \cite{Marolf:2020xie}, the boundary segments were shown to correspond to overlaps $(\psi_j,  \psi_i)$. In our model with spin structures, we can insert an operator reversing the sign of the fermions on the boundary, so we have two kinds of boundary segments, which we will show correspond to $(\psi_j,  \psi_i)$ and $(\psi_j,  (-1)^F  \psi_i)$.   We show that the bulk path integral computes an ensemble average of a product of partition sums, Witten indices, and such inner products where the ensemble includes theories with Hilbert spaces $\mathcal{H}$ of different dimensions, different fractions of fermionic and bosonic states, and randomly chosen gravitational microstates $\psi_i \in \mathcal{H}$.

We then turn to considering ways that the bulk path integral can be modified to make correlation functions of $\hat{\tilde{Z}}$ factorize. In section \ref{alt}, we consider an alternative sum over spin structures where we take odd spin structures with a minus sign. As in \cite{Stanford:2019vob}, this implies that $\hat{\tilde{Z}}$ vanishes as an operator. Thus, factorization is trivially satisfied for $\hat{\tilde{Z}}$; all correlation functions involving $\hat{\tilde{Z}}$ vanish. When we consider the theory with EOW branes, the bulk path integral with boundary segments corresponding to the $(\psi_j, (-1)^F  \psi_i)$ inner product can be non-zero, so long as every boundary includes an even number of such segments. We propose a boundary ensemble interpretation of this alternative path integral where the dimensions of the boundary Hilbert space are doubled, with equal numbers of bosonic and fermionic states. 

In section \ref{single}, we propose a construction of a bulk path integral dual to a single instance of a boundary theory.  An individual theory has a boundary Hilbert space of some fixed dimension $d$, with $m$ fermionic and $d-m$ bosonic states. In \cite{Marolf:2020xie}, the results in a theory with fixed $d$ were reproduced by considering a bulk path integral where the geometry is restricted to have $d$ connected components. We show that correlators of $\hat{\tilde{Z}}$ in a theory with fixed $d$ and $m$ can be reproduced by introducing an additional boundary in each of these connected components with a sum over spin structures on this additional boundary.   Additionally, we analyze coherent ``spacetime D-brane" states, which are special ``boundary conditions" (generalizing the Hartle-Hawking no-boundary condition) for the gravitational path integral that can both modify the boundary ensemble parameters (in a way that cannot be achieved by tuning the bulk couplings) and formally restrict to eigenstates of $\hat{Z}$ or $\hat{\tilde{Z}}$.

In section \ref{jteow}, we consider the extension of these ideas to the sum over spin structures in JT gravity.  We will see that, as in the simple topological model, combining EOW branes with a sum over spin structures is straightforward: a model with a simple sum over all spin structures is dual to the matrix model of \cite{Stanford:2019vob} with EOW branes corresponding to states chosen at random in the boundary Hilbert space as in \cite{Penington:2019kki}. Considering the alternative spin structure is more interesting; we propose a dual description with a single matrix model but a doubled Hilbert space, with the same spectrum of bosons and fermions. We propose an extension of the eigenbrane proposal of  \cite{Blommaert:2019wfy} for describing fixed eigenvalues in the matrix model spectrum by summing over spin structures on the eigenbrane boundary. 

We conclude and summarise directions for future development in section \ref{concl}. The key feature of the examples considered here is that there is a topological obstruction in the bulk for some choices of boundary conditions; this leads to a very simple argument for non-factorization of correlators of boundary creation operators, and thus for an ensemble interpretation in a dual field theory. We discuss the opportunities to find such topological obstructions in more top-down examples of holography in string theory, and directions for investigation of the nature of the baby universe Hilbert space in these models. Finally, we describe an intriguing relation with an older proposal for  describing a time-dependent universe  with an unstable brane placed at  the beginning of time in terms of an ensemble of matrix models of different dimensions.

\section{Topological model with spin structure}
\label{top} 

\subsection{Review of the topological model}

Marolf and Maxfield \cite{Marolf:2020xie} illustrated the general analysis of the path integral in quantum gravity with a simple two-dimensional topological model, where the path integral reduces to a sum over topologies. They considered a path integral over all orientable two-dimensional manifolds with $n$ boundaries. There is no metric or matter field in this model, just smooth topological two-manifolds. The bulk path integral is then a function only of $n$. In anticipation of the boundary ensemble interpretation described in the Introduction, we denote this as $\langle \hat{Z}^n \rangle$ where $\hat{Z}$ is a boundary-creation operator, and the angle brackets are understood in this  context as an expectation in the Hartle-Hawking no-boundary state (see the Introduction for a detailed explanation of the notation). Each connected component, of genus $g$ with $n$ boundaries, is weighted by $e^{\chi S_0+ n S_\partial}$ where $\chi = 2-2g-n$ is the Euler character and $S_0, S_\partial$ are bulk and boundary action parameters. Permutations of compact connected components are treated as a gauge symmetry, introducing a symmetry factor 
\begin{equation}
\mu(M) = \frac{1}{\prod_g m_g!}  
\end{equation}
in the path integral, where $m_g$ is the number of compact connected components of genus $g$.  However, boundaries are treated as distinguishable, so once a component has boundaries there is no analogous symmetry factor. For example, two one-boundary components are not interchangeable because the boundaries are different. In \cite{Marolf:2020xie} the choice $S_\partial = S_0$ is made. Thus,
\begin{equation}
\langle \hat{Z}^n \rangle = \sum_{M : |\partial M| = n} \mu(M) \, e^{S_0 \tilde \chi(M)} ,
\end{equation}
where $\tilde \chi = \sum_{c.c.} (2-2g)$. With this choice, adding boundaries affects the combinatorics but not the weighting. The gravitational path integral with no boundaries is
\begin{equation}
\mathfrak{Z} = \langle 1 \rangle = e^\lambda, 
\end{equation}
where $\lambda$ is the sum over connected manifolds with no boundary, 
\begin{equation}
\lambda = \sum_{g=0}^\infty  e^{S_0 (2-2g)} = \frac{e^{2 S_0} } {1- e^{-2S_0}}. 
\end{equation}
Considering a generating function for  $\hat{Z}^n$,  \cite{Marolf:2020xie} compute 
\begin{equation}
\ln \langle e^{u \hat{Z}} \rangle  = \sum_{connected} \frac{u^n}{n!} e^{S_0 (2-2g)} = \lambda e^u,
\end{equation}
where the sum is over all connected manifolds with arbitrary numbers of boundaries. The result is that 
\begin{equation} \label{bpi}
\mathfrak{Z}^{-1} \langle \hat{Z}^n \rangle = \sum_{d=0}^\infty d^n p_d(\lambda), \quad p_d(\lambda) = e^{-\lambda} \frac{\lambda^d}{d!}.
\end{equation}

There are two ways of interpreting this result for the bulk path integral: the ``boundary" interpretation and the ``baby universe" interpretation. In the boundary interpretation, we consider that each boundary corresponds to a copy of a dual quantum system. Since the boundaries are Euclidean circles, we interpret the path integral for the boundary theory on this circle as dual to the partition function $Z$ in the boundary theory.  (Note again that this is the {\it boundary} path integral and partition function.) Since the bulk has no dynamical fields, the boundary dual is also a  trivial theory with no operators and a vanishing Hamiltonian; we simply have a boundary Hilbert space $\mathcal H$ of dimension $d$. Thus the path integral in a given theory on single boundary computes the partition sum $Z =  \mathrm{tr}_{\mathcal H} (1) = d$.  If there are $n$ boundaries, each carrying a copy of the same theory, the path integral over all of them computes the $n$-fold product $Z^n =  \mathrm{tr}_{\mathcal H} (1) \times \cdots \times \mathrm{tr}_{\mathcal H} (1) $.  Following \cite{Marolf:2020xie}, the bulk path integral is then interpreted as an ensemble average over such theories with different values of $d$. We take an independent boundary Hilbert space of the same dimension for each boundary, and then average over the values of $d$ to get  $\langle Z^n \rangle$, where, in the context of the boundary theory, the angle brackets are understood to mean an ensemble average. The bulk result \eqref{bpi} corresponds to an average where we take each value of $d$ with probability $p_d(\lambda)$. This is just a Poisson distribution for $d$ with mean $\lambda$. 

The baby universe Hilbert space interpretation is obtained by slicing open the bulk path integral along some surface that cuts the bulk manifold into two pieces. The path integral with $n$ boundaries on one side of the slice defines a state $| Z^n \rangle$ on the slice.  As described in the Introduction, we define the Hermitian operator $\hat Z$ which adds a boundary to the path integral on one side of the slice, so $\hat Z |Z^m \rangle = |Z^{m+1} \rangle$. We can then write $|Z^n \rangle = \hat Z^n |\text{HH} \rangle$, where the Hartle-Hawking state $|\text{HH} \rangle = |Z^0 \rangle$ is the state defined by the path integral with no boundaries. The full path integral with $m+n$ boundaries defines an inner product $\langle Z^m| Z^n \rangle$. If the path integral satisfies reflection positivity, this inner product is positive semidefinite, so these states live in a pre-Hilbert space; we can define a baby universe Hilbert space $\mathcal H_{BU}$ by completing the Hilbert space in this norm while quotienting by null states.  

We can  choose a basis of states in $\mathcal H_{BU}$ which are eigenstates of $\hat Z$, $|Z =z \rangle$, where $\hat Z |Z =z \rangle = z |Z=z \rangle$. In the baby universe interpretation, the bulk path integral can be written as the expectation value in the Hartle-Hawking state, $\langle \text{HH} | \hat Z^n | \text{HH} \rangle$. If we expand $|\text{HH} \rangle$ in terms of the $\hat Z$ eigenstates,
\begin{equation} 
|\text{HH} \rangle = \sqrt{\mathfrak{Z}} \sum_z \psi_z |Z = z\rangle, 
\end{equation}
we have
\begin{equation} \label{HHexp}
\langle \text{HH} | \hat Z^n | \text{HH} \rangle = \mathfrak{Z} \sum_z z^n |\psi_z|^2. 
\end{equation}
Comparing to explicit computation of the bulk path integral \eqref{bpi}, we see that the spectrum of $\hat Z$ is $z=d$, $d=0,1, \ldots$, and the amplitudes of the eigenstates in the Hartle-Hawking state are $|\psi_d|^2 = p_d(\lambda)$.

Null states, i.e. states of vanishing norm, appear in this construction because some polynomials in the boundary creation operators vanish identically when acting on physical states (linear combinations of states with definite values of $\dim \mathcal{H}$) \cite{Marolf:2020xie}. The appearance of these states is associated with a generalised diffeomorphism symmetry: the slicing of the bulk is completely characterised by the number of asymptotic boundaries to one side of it, so that bulk slices which differ in the interior but have the same boundaries are regarded  as equivalent. The restriction of the spectrum of $\hat Z$ to discrete values implies that, given any function $f(x)$ which vanishes for $x= 0,1, \ldots$, the state $|f(Z) \rangle = f(\hat Z) |\text{HH} \rangle$ in the pre-Hilbert space is a null state; that is, $f(\hat Z)$ annihilates all physical states in $\mathcal H_{BU}$. Thus $f(\hat{Z})$ acting on the Hartle-Hawking vacuum formally creates a state of vanishing norm, i.e., a null state.

In this example, the gravitational path integral does not factorize between the spacetime boundaries, i.e., $\langle \hat{Z}^n \rangle \neq \langle \hat{Z} \rangle^n$, as is evident from the sum on the RHS of \eqref{bpi}.  The form of  \eqref{bpi} suggests a dual boundary interpretation in terms of an ensemble average , and a bulk interpretation in terms of the expansion of the Hartle-Hawking state in $\hat Z$-eigenstates in the baby universe picture. The general point of \cite{Marolf:2020xie} is that any quantum gravity path integral will have a non-trivial expansion like \eqref{HHexp}, and the bulk path integral will not factorize,  unless $|\text{HH} \rangle$ is an eigenstate of the operators $\hat Z[J]$. The standard bulk path integral cannot then not be dual to a single boundary theory, and at best can have an boundary ensemble interpretation.\footnote{McNamara and Vafa have conjectured \cite{McNamara:2020uza} that in every consistent theory of quantum gravity in $d>3$ dimensions, the baby universe Hilbert space is one-dimensional (the quotient by the null states identifies all states in the pre-Hilbert space), so that such ensemble interpretations do not arise.} 

The baby universe picture does suggest alternative quantities that may factorise; instead of taking the expectation value of $\hat Z^n$ in the Hartle-Hawking state, we can take it in an eigenstate, $\langle Z=d | \hat Z^n | Z=d \rangle = d^n$. This corresponds to evaluating the bulk path integral with $n$ boundary creation operators inserted between two slices each carrying the baby universe state with $Z=d$. It is not clear  how to do this in general, but, remarkably, in the simple topological model it corresponds to restricting the path integral to manifolds with precisely $d$ connected components \cite{Marolf:2020xie}.   In the dual theory it corresponds to calculating correlators in a fixed field theory where each boundary carries a $d$-dimensional Hilbert space, rather than an ensemble of theories.

\subsection{Adding spin structure}

\begin{figure}[t]
    \centering
    \includegraphics[scale=.1]{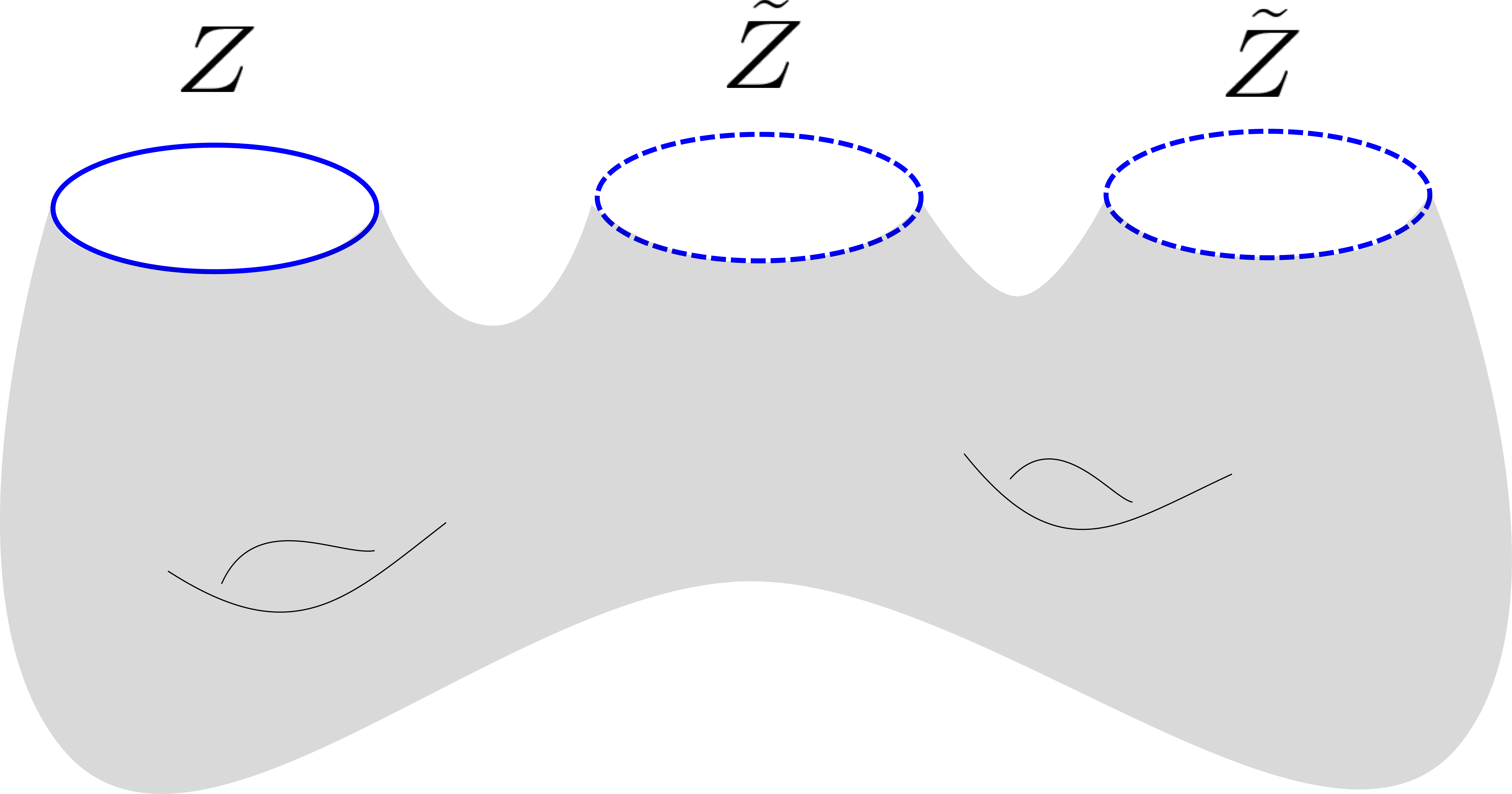}
    \caption{\small{A genus $g=2$ contribution to the bulk path integral with one NS and two R boundaries.  With these boundary conditions the gravitational path integral computes $\avg{\hat{Z} \hat{\tilde{Z}}^2}$.  The depicted diagram's contribution to this expectation is $64e^{-5S_0+3S_\partial}$. Asymptotic boundaries with NS (R) spin structure are shown as solid (dashed) blue circles.  In the free energy \eqref{eq:ramond-free-energy}, this connected manifold contributes $32e^{-5S_0+3S_\partial}$ to the coefficient of $u\tilde{u}^2$ (the only difference compared to its contribution to $\avg{\hat{Z}\hat{\tilde{Z}}^2}$ is the additional symmetry factor $(n_{NS}! n_R!)^{-1}$ which arises in the usual way by passing from the generating function to the free energy).  In the alternative sum over spin structures discussed in section \ref{alt}, the contribution from this manifold to both $\avg{\hat{Z} \hat{\tilde{Z}}^2}$ and the free energy \eqref{eq:alt-ramond-free-energy} vanishes identically since $n_R \neq 0$.}}
    \label{fig:ramond}
\end{figure}

We aim to further illustrate these ideas by extending the topological model above by equipping the bulk spacetimes with a spin structure, and summing over the spin structures in the bulk consistent with a choice of spin structure on the asymptotic boundaries. In this section, we will calculate the bulk path integral, show that it also has a natural ensemble interpretation in the boundary, and comment on the baby universe interpretation. 

As explained in the Introduction, we will have two types of boundaries, corresponding to NS and R spin structure, created by $\hat{Z}$ and $\hat{\tilde Z}$ respectively. The dual field theory on these boundaries inherits the NS or R boundary conditions for fermions, and thus the  path integral on one of these boundaries computes $\tr_{\mathcal H}(1)$ or $\tr_{\mathcal H}((-1)^F)$ respectively. The  gravitational path integral with a given set of boundaries computes the expectation value $\langle \hat{Z}^{n_{NS}} \hat{\tilde{Z}}^{n_R} \rangle$ in the Hartle-Hawking vacuum, and involves a sum over all spin two-manifolds with the given boundaries (see an example in figure \ref{fig:ramond}).  We take the action for a connected component in the bulk to be $\chi S_0 + n S_\partial$, where $\chi$ is the Euler character and $n$ is the number of boundaries, independently of the choice of spin structure.  Thus, the sum over spin structures in the path integral just multiplies the contribution of each connected component by a factor counting the number of distinct spin structures consistent with the boundary conditions. 

Stanford and Witten \cite{Stanford:2019vob} show that for a genus $g$ two-manifold with $n$ boundaries, of which $n_R$ have Ramond spin structure, the sum over spin structures gives
\begin{equation}
\sum_{spin} 1 = 2^{2g+n-2} (1+(-1)^{n_R}) \,.
\end{equation}
Using this formula we get
\begin{equation}
\ln \langle e^{u \hat{Z} + \tilde u \hat{\tilde Z}} \rangle = \sum_{g, n_{NS}, n_R} e^{S_0(2-2g-n)} e^{n S_\partial} \frac{u^{n_{NS}}}{n_{NS}!} \frac{\tilde u^{n_R}}{n_R!} 2^{2g+n-2} (1+(-1)^{n_R}), \label{eq:ramond-free-energy} 
\end{equation}
where the sum is over all connected geometries. In other words, we sum over genus $g$ and the number of NS and R boundaries,  $n_{NS}, n_R$, where $n = n_{NS} + n_R$ is the total number of boundaries. We have already summed over spin structures for each geometry to produce the final factor. In this case, it is convenient to take $S_0 = \tilde S_0 + \ln 2 $ and $S_\partial = \tilde S_0$,\footnote{As in \cite{Marolf:2020xie}, adding a constant to $S_\partial$ will scale the eigenvalues of $\hat{Z}$.}  so that the sum splits into  separate factors, 
\begin{equation}
\ln \langle e^{u \hat{Z} + \tilde u \hat{\tilde Z}} \rangle = \sum_{g} e^{\tilde S_0(2-2g)}  \sum_{n_{NS}}  \frac{u^{n_{NS}}}{n_{NS}!}  \sum_{n_R} \frac{\tilde u^{n_R}}{n_R!} (1+(-1)^{n_R}) = \lambda e^u \cosh(\tilde u),
\label{eq:ren-free-energy}
\end{equation}
where
\begin{equation}
\ln \mathfrak{Z} = \lambda = 2 \sum_g e^{\tilde S_0(2-2g)} = \frac{2 e^{2\tilde S_0}}{1-e^{-2 \tilde S_0}}.    
\end{equation}
We can expand this out as:
\begin{equation}
\mathfrak{Z}^{-1} \langle e^{u \hat{Z} + \tilde u \hat{\tilde Z}} \rangle = \sum_{d=0}^\infty p_d(\lambda) e^{ud} \sum_{m=0}^d \left( \begin{array}{c} d \\ m \end{array} \right) 2^{-d} e^{\tilde u (d-2m)},  
\end{equation}
where $p_d(\lambda) = e^{-\lambda} \frac{\lambda^d}{d!}$ as before.
Extracting the coefficient of $\frac{u^{n_{NS}}}{n_{NS}!} \frac{\tilde u^{n_R}}{n_R!}$, we have 
\begin{equation}
\mathfrak{Z}^{-1} \langle \hat{Z}^{n_{NS}} \hat{\tilde{Z}}^{{n_R}} \rangle = \sum_{d=0}^\infty p_d(\lambda) d^{n_{NS}} \sum_{m=0}^d \left( \begin{array}{c} d \\ m \end{array} \right) 2^{-d} (d-2m)^{n_R}.
\label{spinresult}
\end{equation}
Note that $\langle \hat{\tilde Z} \rangle = 0$ is zero. Indeed the RHS  vanishes for all odd $n_R$, analogously to the discussion in \cite{Stanford:2019vob}.

Our first result is that this gravitational path integral has a natural interpretation in an ensemble of boundary theories.  Consider $n$ copies of a field theory with a $d$-dimensional Hilbert space and vanishing Hamiltonian on a circle.  Let $n_{NS}$ of the circles  have anti-periodic fermions, and $n_R$ have periodic fermions.  The path integral on these $n$ circles computes $Z^{n_{NS}} \tilde{Z}^{n_R} = \left[\tr_{\mathcal H}(1))\right]^{n_{NS}} \left[\tr_{\mathcal H}((-1)^F)\right]^{n_R}$.  The first sum in \eqref{spinresult} can be interpreted as before, as an ensemble average where $Z$ takes non-negative integer values $d$ with a Poisson distribution with mean $\lambda$.  Thus we have a boundary ensemble in which we sum over the Hilbert space dimension with Poisson weight.  The second sum can be interpreted by requiring that $\tilde Z$ take values in $(-d, \ldots d)$ by steps of 2 with binomial probabilities. We can reproduce this from a boundary description if we consider a Hilbert space of dimension $d$, with $m$ fermionic and $d-m$ bosonic states, so $Z = \tr_{\mathcal H}(1) = d$ and $\tilde Z = \tr_{\mathcal H}((-1)^F) = d-2m$. The binomial probabilities can be obtained if we treat the $d$ states in the Hilbert space as distinguishable and take each to be bosonic or fermionic with equal probability.\footnote{Equivalently, to describe the $Z$ and $\tilde{Z}$ distributions compactly together, we could say that we draw the number of bosonic states $d_b$ and the number of fermionic states $d_f$ independently at random with Poisson statistics each having mean $\lambda/2$.}  With this identification we can relate correlation functions of boundary creation operators in the baby universe Hilbert space to a dual ensemble averaged field theory: $ \langle \hat{Z}^{n_{NS}} \hat{{\tilde Z}}^{n_{R}} \rangle = \langle {Z}^{n_{NS}} {\tilde Z}^{n_{R}} \rangle $ where the angle brackets on the left side refer to expectation in the Hartle-Hawking no-boundary vacuum, and the angle brackets on the right side refer to dual ensemble average.

The correlation functions of $\tilde Z$ provide a particularly nice manifestation of the failure of factorization associated with this ensemble interpretation. There are no bulk spin structures compatible with an odd number of Ramond ($\tilde{Z}$) boundaries, so the gravitational path integral with this boundary condition vanishes because there is nothing to integrate over, e.g,, $\langle \tilde Z \rangle = \langle \hat{\tilde{Z}} \rangle = 0$. However, the gravitational path integral with  an even number of Ramond boundaries is non-zero, so it clearly does not factorize, e.g., $\mathfrak Z^{-1} \langle \tilde Z^2 \rangle \neq \mathfrak Z^{-2} \langle \tilde Z \rangle^2$.

The sum over spin structures also provides an interesting illustration of the appearance of null states in the baby universe Hilbert space which is  spanned by the simultaneous eigenstates of $\hat{Z}, \hat{\tilde Z}$. Since the spectrum is restricted, there are combinations of the operators $\hat{Z}, \hat{\tilde Z}$ which annihilate all the states in the baby universe Hilbert space. 
This was explained in \cite{Marolf:2020xie} for the discrete spectrum of $\hat{Z}$; the case of $\hat{\tilde Z}$ provides an even simpler example.   In parallel with \eqref{HHexp}, we want to expand the Hartle-Hawking vacuum in simultaneous eigenstates of $\hat{Z}$ and $\hat{\tilde{Z}}$, $|\text{HH}\rangle = \sqrt{\mathfrak{Z}} \sum_{z,\tilde{z}} \psi_{z\tilde{z}} |Z=z, \tilde{Z} = \tilde{z} \rangle$, which gives the correlation $\langle \text{HH} | \hat{Z}^{n_{NS}} \hat{\tilde{Z}}^{n_R} | \text{HH} \rangle = \mathfrak{Z} \sum_{z,\tilde{Z}} |\psi_{z,\tilde{z}}|^2$.  We can read off the possible values of $z$ and $\tilde{z}$ by comparison with  the explicit computation of the gravitational path integral in \eqref{spinresult}. The outer sum  shows that $z=0,1,2,\cdots$ is the sum over the spectrum of $\hat{Z}$, and the inner sum shows that $\tilde{z} = -d , -d +2, \cdots d$ is the spectrum of $\hat{\tilde Z}$. The spectrum of $\hat{\tilde Z}$ depends on the spectrum of $\hat{Z}$. So, to characterise the null states associated with $\hat{\tilde Z}$, we must split the baby universe Hilbert space into a direct sum of eigenspaces of $\hat{Z}$, and in each eigenspace consider appropriate operators. In the eigenspace with $z=1$, the possible eigenvalues of $\hat{\tilde{Z}}$ are $\tilde{z} = \pm 1$, so $\hat{\tilde Z}^2-1$ vanishes as an operator on this subspace, and indeed any operator containing a factor of $\hat{\tilde Z}^2 -1$, such as $\hat{\tilde Z}^3- \hat{\tilde Z}$, will  vanish as an operator on this subspace. Similarly, in the eigenspace where $z=2$, the possible values are $\tilde z = -2, 0, 2$, so all polynomials with a factor of $\hat{\tilde Z} (\hat{\tilde Z}^2 -4)$ vanish as operators on this subspace.\footnote{In the language of \cite{Gesteau:2020wrk}, the addition of spin structure enlarges the group of baby universe transformations from $\mathbb{Z}$ to $\mathbb{Z}^2$, but the GNS representation of $C^*(\mathbb{Z}^2)$ induced by the state $\omega$ (which defines the gravitational path integral) is not irreducible. 
We further point out that there is an intriguing interplay between the choice of $\omega$, the group of baby universe transformations, and the quotient by null states which produces the GNS Hilbert space.
In particular, the commutative nature of $C^*(\mathbb{Z}^2)$ does not necessarily imply that the eigenvalues of the group generators (acting on the GNS Hilbert space) are independent of each other.  The allowed eigenvalues can instead become coupled by the choice of $\omega$, and this is what occurs in our discussion of spin structures.} 

\section{End of the world branes}
\label{eow}

We can extend this model by considering end of the world (EOW) branes \cite{Kourkoulou:2017zaj,Penington:2019kki}, which were adapted to the topological theory in \cite{Marolf:2020xie}. 
As discussed in the Introduction, for the theory with EOW branes, the boundary conditions include circular boundaries with either NS or R spin structure and boundary segments, with EOW branes intersecting the boundary at each end of the segment. The EOW branes carry flavour labels $i = 1, \ldots k$, and specifying a boundary segment will fix the flavour labels at the  endpoints. In the theory with spin structure the segments also carry an even/odd index. In the calculation of the bulk path integral, we sum over all ways of pairing the EOW branes starting and ending on the boundary segments, sewing the boundary segments into circular boundaries of the bulk two-manifold. The state index on either end of an EOW brane must be identical. If the resulting circle boundary contains an even number of odd-index segments, it has an NS spin structure, otherwise it has an R spin structure. We also allow pure EOW brane boundaries -- circular branes on which the bulk geometry is allowed to terminate. These pure EOW brane boundaries are not part of the asymptotic boundary data which defines the correlator, and we must sum over all the ways they may be inserted, just as we sum over all possible bulk topologies. We will assume that these pure EOW brane boundaries have an NS spin structure. This is based on the intuition that in a theory with more dynamics, there would be a limit where the EOW brane shrinks to zero size, and we would like a zero size EOW brane boundary to be equivalent to no boundary.  We  then sum over all bulk geometries and spin structures with a given set of circular boundaries.

The boundary dual field theory still has factors that are defined on the $Z$ and $\tilde{Z}$ on which the path integral with odd and even boundary conditions for fermions computes the partition sum and the Witten index respectively.  But additionally there are factors defined on the boundary segments, again with a $d$-dimensional Hilbert space ${\cal H} = {\cal H}_b \oplus {\cal H}_f$ with bosonic and fermionic components of dimension $d_b$ and $d_f$ respectively, and a vanishing Hamiltonian. Now  a segment of the asymptotic boundary with an EOW brane and index $i$ at one end is interpreted as defining a state $\psi_i \in {\cal H}$ at the free end.
We can also consider the path integral 
on a segment of asymptotic boundary ending on EOW branes with labels $i, j$ at the two ends. For a boundary with an even index we interpret this path integral as
a standard inner product in the boundary Hilbert space $(\psi_j, \psi_i)$; we will call this a ``standard'' (or NS) segment.
For a boundary with an odd index we interpret the path integral on a segment as computing  the twisted inner product $(\psi_j, (-1)^F \psi_i)$ instead; we will call this a ``twisted'' (or R) segment.
In detail, if $\ket{\psi_i} = \sum_{m=1}^{d_b} \psi_i^m |b_m\rangle + \sum_{n=1}^{d_f} \psi_i^n |f_n\rangle$ where $|b_m\rangle$ and $|f_n\rangle$ are  bases for ${\cal H}_b$ and ${\cal H}_f$, then $(-1)^F\ket{\psi_i} = \sum_{m=1}^{d_b} \psi_i^m |b_m\rangle - \sum_{n=1}^{d_f} \psi_i^n |f_n\rangle$.  In other words, $(-1)^F$ flips the sign of the fermionic components of a state.

In  \cite{Marolf:2020xie}, following \cite{Penington:2019kki}, the path integral with just the $(\psi_j, \psi_i)$ segments was shown to have an ensemble interpretation where the states $\psi_i$ are randomly chosen in the boundary Hilbert space: for each choice of boundary Hilbert space dimension $d$, the components $\psi_i^a$ in some basis $\{ |a \rangle \}$ (where $a = 1, \ldots d$) were i.i.d. Gaussian random variables.  We will show that there is a similar ensemble interpretation in our case, with the addition that as in the previous section, we must  take the basis states for the Hilbert space to be randomly bosonic or fermionic with equal probability.

Let $\hat{Z}$ and $\hat{\tilde{Z}}$ create boundaries with NS and R boundary conditions as in the previous sections, and $\hat{S}_{ji}$ and $\hat{\tilde{S}}_{ji}$ create even and odd boundary segments with labels $j$ and $i$ on the two endpoints. We consider the generating function 
\begin{equation}
\langle e^{u \hat{Z} + \tilde u \hat{\tilde{Z}} + t_{ij} \hat{S}_{ji}+ \tilde t_{ij} \hat{\tilde{S}}_{ji}}  \rangle ,
\end{equation}
which involves a sum over all numbers of asymptotic boundaries and boundary segments.   The angle brackets indicate an expectation value in the Hartle-Hawking no-boundary vacuum.  In this generating function, a contribution with an asymptotic boundary of type $Z$, $\tilde Z$ comes with a factor of $u$, $\tilde u$, and a circular asymptotic boundary formed by summing together boundary segments comes with a factor of the trace of the matrix product of the corresponding chemical potentials $t_{ij}$, $\tilde t_{ij}$.  Note that the repeated $i,j$ indices in the exponent are summed over.

\begin{figure}[t]
    \centering
    \includegraphics[scale=.1]{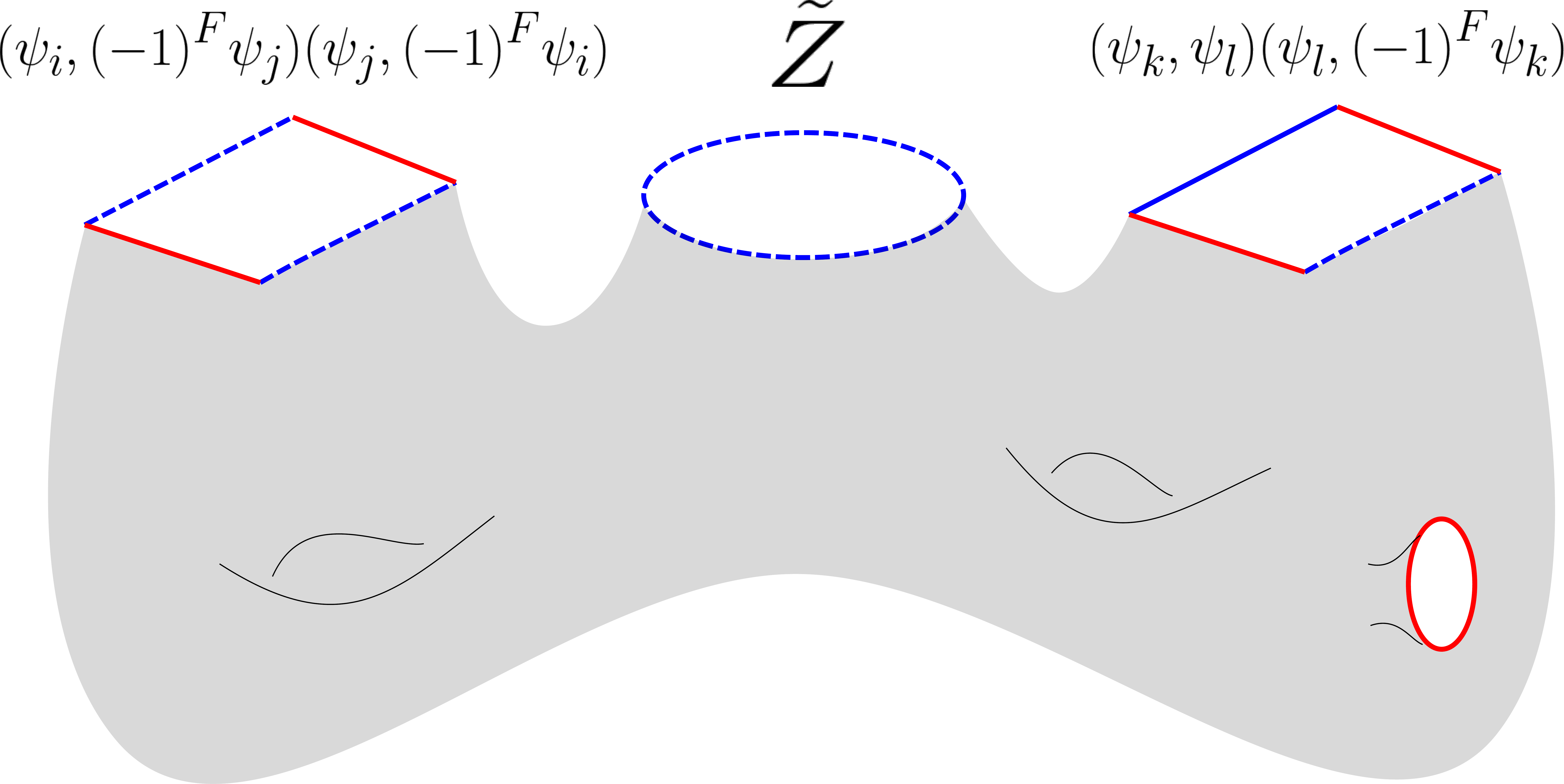}
    \caption{\small{A contribution to the expectation $\avg{\hat{\tilde{S}}_{ij} \hat{\tilde{S}}_{ji} \hat{\tilde{Z}} \hat{S}_{kl} \hat{\tilde{S}}_{lk}}$.  We do not sum over indices in such expressions; if we had not matched the outgoing EOW brane label of one boundary segment with the incoming EOW brane label of the subsequent boundary segment, the bulk path integral on this manifold would have been zero.  Red segments are EOW branes and solid (dashed) blue segments are asymptotic NS (R) boundary segments.  We can of course still have circular NS or R boundaries as before.  There can also be closed EOW branes (red circle) floating in the bulk, which are not part of the boundary data. }}
    \label{fig:eow-topological}
\end{figure}

The bulk path integral treats all types of NS or R boundaries in the same way, so the generating function is a function of the ``total" NS and R chemical potentials, 
\begin{equation}
U = u + T_{even} = u +  \sum_{n=1}^\infty \frac{1}{n} \mathrm{tr}(t^n) + \sum_{n =0, n \neq m}^\infty \sum_{m=0}^\infty  \mathrm{tr} (t^n \tilde t t^m \tilde t) + \frac{1}{2} \sum_{n=0}^\infty \mathrm{tr} ( t^n \tilde t t^n \tilde t) +  \ldots,
\end{equation}
\begin{equation}
\tilde U = \tilde u + T_{odd} = \tilde u + \sum_{n=0}^\infty \mathrm{tr}( t^n \tilde t)   + \ldots . 
\end{equation}
where $T_{even}$ is the sum of all the traces with an even number of $\tilde t_{ij}$ factors, and $T_{odd}$ is the sum of all traces with an odd number of $\tilde t_{ij}$ factors.\footnote{We include a symmetry factor in cases where the trace is invariant under some subset of cyclic permutations.} We make the same choice as before that $S_0 = \tilde S_0 + \ln 2$, $S_\partial = \tilde S_0$. The sum over connected manifolds with no asymptotic boundaries includes a sum over an arbitrary number of pure EOW brane boundaries (``floating" in the bulk as in figure \ref{fig:eow-topological}), which introduce a factor of $k$ from the trace over the EOW degrees of freedom, so 
\begin{equation}
\ln \mathfrak Z = \ln \langle 1 \rangle =\lambda = \frac{2 e^{2 \tilde S_0} } {1- e^{-2\tilde S_0}} e^k. 
\end{equation}
The bulk path integral then gives 
\begin{equation}
\langle e^{u \hat{Z} + \tilde u \hat{\tilde{Z}} + t_{ij} \hat{S}_{ji} + \tilde t_{ij} \hat{\tilde{S}}_{ji}} \rangle = e^{\lambda e^U \cosh \tilde U} \,,
\end{equation}
where the repeated $i,j$ indices in the exponent are summed.
Expanding this out, we have 
\begin{equation}
\mathfrak{Z}^{-1} \langle e^{u \hat{Z} + \tilde u \hat{\tilde{Z}} + t_{ij} \hat{S}_{ji} + \tilde t_{ij} \hat{\tilde{S}}_{ji} } \rangle = \sum_{d=0}^\infty p_d(\lambda) e^{Ud} \sum_{m=0}^d \left( \begin{array}{c} d \\ m \end{array} \right) 2^{-d} e^{\tilde U (d-2m)},  
\label{eowresult}
\end{equation}

We want to equate the correlation functions of boundary creation operators in the baby universe Hilbert space $\langle \hat{Z}^m \hat{\tilde{Z}}^{\tilde{m}} \hat{S}_{ji} \cdots \hat{\tilde{S}}_{kl} \cdots \rangle$ with an ensemble average over  boundary theories $\langle Z^m \tilde{Z}^{\tilde{m}} (\psi_j,\psi_i) \cdots (\psi_k, (-1)^F \psi_l) \rangle$, where $Z$ is the partition sum, $\tilde{Z}$ is the Witten index, and the sequence of inner products and twisted inner products arises from the segments ending on EOW branes.  As in the previous section, we can interpret the sum over $d$ in \eqref{eowresult} as choosing the dimension $d$ of the boundary Hilbert space leading to the partition sum $Z=d$, and the sum over $m$ as determining the number of fermionic states in this Hilbert space, leading to the Witten index $\tilde Z = d-2m$. Fixing these quantities, we would like to interpret the remaining dependence on $t_{ij}$, $\tilde t_{ij}$ as defining a boundary ensemble average as follows:
\begin{equation} \label{dmeow} 
\langle e^{t_{ij} (\psi_j, \psi_i) + \tilde t_{ij} (\psi_j, (-1)^F \psi_i)} \rangle_{d,m} = e^{d T_{even} + (d-2m) T_{odd}}.
\end{equation}

We want to determine the correct  ensemble for the average in \eqref{dmeow}. In \cite{Marolf:2020xie}, the corresponding factor was 
\begin{equation} 
e^{d \sum_{n=1}^\infty \frac{1}{n} \mathrm{tr} (t^n)} = \frac{1}{\det(I-t)^d},
\end{equation}
which was recognised as the generating function of a complex Wishart distribution, which can be written as a Gaussian integral in terms of random variables $\psi^a_i$, 
\begin{equation} 
 \frac{1}{\det(I-t)^d} = \frac{1}{\pi^{kd}}\int \prod_{i=1}^k \prod_{a=1}^d   d\psi^a_i d\bar \psi^a_i e^{-\bar \psi^a_i \psi^a_i} \exp \left( t_{ij} \sum_{a=1}^d \bar \psi^a_j \psi^a_i \right).
\end{equation}
In this expression,  $\psi_i^a$ was interpreted as the coefficient in a basis expansion of states: $|\psi_i\rangle = \sum_{a=1}^d \psi_i^a \ket{a}$.
To reproduce \eqref{dmeow}, we just need a simple generalisation of this Gaussian integral,
\begin{equation} \label{gint}
\begin{split}
 e^{d T_{even} + (d-2m) T_{odd}} = \frac{1}{\pi^{kd}}\int & \prod_{i=1}^k \prod_{a=1}^d   d\psi^a_i d\bar \psi^a_i  e^{-\bar \psi^a_i \psi^a_i} \\
 & \times \exp \left[ t_{ij} \sum_{a=1}^d \bar \psi^a_j \psi^a_i + \tilde t_{ij} \left( - \sum_{a=1}^{m} \bar \psi^a_j \psi^a_i + \sum_{a=m+1}^d \bar \psi^a_j \psi^a_i \right) \right].
 \end{split}
\end{equation}
To see this, expand the exponential in $t, \tilde t$ in a Taylor series.
In each term, performing the Gaussian integral over $\psi, \bar \psi$ will result in Wick contractions between the different $\psi$'s, forming products of traces of $t$, $\tilde t$. For each trace, there is a single remaining sum over $a$. In a term with an even number of $\tilde t$ in a trace, all the different $\psi^a$ will contribute in the same way, giving an overall factor of $d$. With an odd number of $\tilde t$, the values $a=1, \ldots m$ will contribute with a minus sign, giving an overall factor of $d-2m$. 

The interpretation of this is that the $|\psi_i\rangle$ are simply random states in the Hilbert space, as before, and are random linear combinations of the bosonic and fermionic basis elements. In $(\psi_j, (-1)^F \psi_i)$, the fermionic elements, which we have taken for definiteness to be labeled by $a=1,\ldots m$, contribute with the opposite sign. 

The addition of the twisted boundary segment creation operators $\hat{\tilde{S}}_{ji}$ corresponding to $(\psi_j, (-1)^F \psi_i)$  enlarges the Hilbert space of baby universes, just as the addition of the Ramond boundary creation operators  $\hat{\tilde{Z}}$ corresponding to Witten indices $\tilde{Z}$ did in the previous section. In \cite{Marolf:2020xie}, the baby universe Hilbert space for fixed $d$ was $L^2(M_k^d)$, where $M_k^d$ is the space of Hermitian positive definite $k \times k$ matrices with rank $\leq d$, corresponding to the inner product matrices $(\psi_j,  \psi_i)$ that can be constructed from $\psi_i^a$ with $a = 1, \ldots d$. The dimension of $M_k^d$ is $k^2$ for $k \leq d$ and $ 2kd-d^2$ for $k \geq d$. For the present case, it is convenient to combine our boundary observables to form the matrices $(\psi_j,  (1 \pm (-1)^F )\psi_i)$, which give the inner product between the restrictions of our states to the bosonic or fermionic subspaces respectively. Thus the baby universe Hilbert space for fixed $d, m$ is $L^2(M_b) \otimes L^2(M_f)$, where $M_b = M_k^{d-m}$ and $M_f = M_k^m$ are the  space of Hermitian positive definite $k \times k$ matrices with rank $\leq (d-m)$ and rank $\leq m$, respectively.
The parameter space $M_b \oplus M_f$ then has dimension
\begin{equation}
\dim (M_b \oplus M_f) = 
\begin{cases}
    2dk-m^2-(d-m)^2, & k > d-m,m\\
    k^2+2km-m^2, & d-m \geq k > m \\
    k^2+2k(d-m)-(d-m)^2, & m \geq k >d-m \\
    2k^2, & d-m,m \geq k .
\end{cases}
\end{equation}

\section{Alternative sum over spin structures}
\label{alt}

So far we have summed with equal weight over all spin structures consistent with a given boundary condition.   
We can alternatively weight the sum over spin structures by $(-1)^\zeta$, where $\zeta$ is the number of zero modes (mod 2) of the Dirac equation $\slashed{D} \lambda = 0$ with the given choice of spin structure,\footnote{For a fixed choice of spin structure, $\zeta$ is anomalous on manifolds with boundary if the spin structure is Ramond on some of the boundaries \cite{Dijkgraaf:2018vnm}. However, the variation in $\zeta$ as we change the spin structure in the bulk while holding boundary spin structure fixed is well-defined, so the sum of $(-1)^\zeta$ over spin structures with a given boundary spin structure is well-defined up to an overall sign for cases with some Ramond boundaries. As the value vanishes in these cases, the overall sign ambiguity is unimportant.} a choice that divides  spin structures into even and odd classes. With this choice, \cite{Stanford:2019vob} found that the sum over spin structures gives 
\begin{equation} 
\sum_{\mathrm{spin}} (-1)^\zeta = \left\{ \begin{array}{ll}  2^{g+n-1} &  \mathrm{if }\  n_R=0 \\ 0 & \mathrm{otherwise.} \end{array} \right. 
\end{equation}
Thus, the answer is zero for all cases with Ramond boundaries (e.g., the case in figure \ref{fig:ramond}). Without EOW branes, the model is then very similar to the one in \cite{Marolf:2020xie}. The sum over connected geometries is 
\begin{equation}
\ln \langle e^{u \hat{Z} + \tilde u \hat{\tilde{Z}}} \rangle = \sum_{g, n_{NS}, n_R} e^{S_0(2-2g-n)} e^{n S_\partial} \frac{u^{n_{NS}}}{n_{NS}!} \frac{\tilde u^{n_R}}{n_R!} 2^{g+n-1} \delta_{n_R,0} = \sum_{g, n} e^{S_0(2-2g-n)} e^{n S_\partial} \frac{u^{n}}{n!} 2^{g+n-1} 
\label{eq:alt-ramond-free-energy}
\end{equation}
Thus $\hat{\tilde{Z}}$ is zero as an operator, as the generating function is independent of $\tilde u$. We can absorb the factor of 2 in $S_0$ by taking $S_0 = \tilde S_0 + \frac{1}{2} \ln 2$, and make the $n$ dependence cancel out by taking $S_\partial = \tilde S_0 - \frac{1}{2} \ln 2$. Then 
\begin{equation}
\ln \mathfrak Z = \lambda = \frac{ e^{2 \tilde S_0} } {1- e^{-2\tilde S_0}} 
\end{equation}
and
\begin{equation}
\mathfrak{Z}^{-1} \langle e^{u \hat{Z} + \tilde u \hat{\tilde Z}} \rangle = e^{-\lambda} e^{\lambda e^u} = \sum_{d=0}^\infty p_d(\lambda) e^{ud}.
\end{equation}
We have an ensemble labelled by a single integer $d$, which could be interpreted as the dimension of the boundary Hilbert space. Thus correlators of $\hat{\tilde{Z}}$ factorize trivially since they all vanish.

The structure is more interesting if we include EOW branes. The twisted boundary segments $(\psi_j, (-1)^F \psi_i) $ still have a non-trivial effect, as  boundaries with an even number of these factors have NS boundary conditions. Thus, with EOW branes one can show that
\begin{equation}
\ln \mathfrak Z = \lambda = \frac{e^{2 \tilde S_0} } {1- e^{-2\tilde S_0}} e^k,
\end{equation}
and
\begin{equation}
\langle e^{u \hat{Z} + \tilde u \hat{\tilde Z} + t_{ij} 
\hat{S}_{ji}
+ \tilde t_{ij}
\hat{\tilde{S}}_{ji}}
\rangle = e^{\lambda e^U},
\end{equation}
where $U = u + T_{even}$ as before,  $\hat{S}_{ji}$ creates standard $(\psi_j, \psi_i)$ segments, and $\hat{\tilde{S}}_{ji}$ creates twisted $(\psi_j, (-1)^F \psi_i)$ segments. Recall that $T_{even}$ depends on both $t_{ij}$ and $\tilde t_{ij}$. Conditioning on  eigenstates of $\hat{Z}$ with eigenvalue $Z=d$,
\begin{equation} 
\langle e^{t_{ij} \hat{S}_{ji}+ \tilde t_{ij} \hat{\tilde{S}}_{ji}} \rangle_{d} = e^{d T_{even}}.    
\end{equation}
To provide a boundary description of this result, we must give the boundary Hilbert space additional structure corresponding to the difference between the standard $(\psi_j, \psi_i)$ and  twisted $(\psi_j, (-1)^F \psi_i)$ segments. A natural interpretation is that the boundary system is as in the previous section, but is restricted to have an equal number of fermionic and bosonic states. This might seem problematic, as the total Hilbert space dimension $d$ could be even or odd, but we can take  the boundary Hilbert space to consist of $d$ bosonic and $d$ fermionic dimensions, with the $\psi_i^a$ coefficients of the basis states chosen as i.i.d. Gaussian random variables. This makes the Witten index $\tilde Z$ and combinations with an odd number of factors of $(\psi_j, (-1)^F \psi_i)$ vanish as desired in the boundary ensemble. However, it means that the partition sum in a given element of the ensemble is $Z  = \mathrm{tr}_{\mathcal H}(1) = 2d$, rather than $d$, and similarly for the non-vanishing combinations of boundaries with EOW branes. We can deal with this either by instead identifying $Z = \frac{1}{2} \mathrm{tr}_{\mathcal H}(1)$ (similar to the choice in \cite{Stanford:2019vob} that $Z_{NS}(\beta) = \sqrt{2} \tr e^{-\beta H}$), or by changing our choice for $S_\partial$: if $S_\partial =  S_0 = \tilde S_0 + \frac{1}{2} \ln  2$, we find that 
\begin{equation}
\mathfrak{Z}^{-1} \langle e^{u \hat{Z} + \tilde u \hat{\tilde{Z}}+ t_{ij} 
\hat{S}_{ji}
+ \tilde{t}_{ij} 
\hat{\tilde{S}}_{ji}
}
\rangle 
= \sum_{d=0}^\infty p_d(\lambda) e^{2Ud},  
\end{equation}
consistent with a boundary interpretation with $Z = \mathrm{tr}_{\mathcal H}(1)$ taking even integer values. 

Thus, once we include the EOW branes we see that the natural boundary interpretation of our bulk path integral has more structure, and we propose that the theory where we sum over spin structures weighted by $(-1)^\zeta$ is dual to an ensemble of boundary theories which each have $d$ bosonic and $d$ fermionic states.

\section{Dual of a single theory}
\label{single}

We would like to identify a bulk dual description for the individual theories of fixed Hilbert space dimension, where we fix the numbers of bosonic and fermionic states (not necessarily equal), to make progress towards linking these ideas to the conventional AdS/CFT picture, where we have a bulk gravity theory dual to a specific boundary theory, rather than an ensemble of theories. In the baby universe picture, we are looking for a bulk calculation that gives the expectation values in an eigenstate of the boundary creation operators $|Z = d, \tilde Z = d-2m \rangle$ rather than the Hartle-Hawking state. In \cite{Marolf:2020xie}, the theory with a specific dimension $d$ for the boundary Hilbert space was identified with the gravitational path integral restricted to the sector where the bulk has precisely $d$ connected components.\footnote{See also \cite{Giddings:2020yes} for an attempt to reconcile the ensemble average induced by wormholes with the unitarity of the underlying quantum mechanical evolution.} 
The idea is basically that if we have $d$ connected components, when we add an additional boundary we get to choose which component we add it to, so the bulk path integral gets multiplied by a factor of $d$. 

We would like to give a similar description for the eigenstates of $\hat{\tilde{Z}}$.  The gravitational path integral that we described vanishes if there is a single $\tilde{Z}$ boundary because there is no bulk manifold with spin structure consistent with this boundary condition.  However, in the dual description, if we have a  single theory with  fixed but unequal numbers of bosonic and fermionic states, the Witten index $\tilde{Z}$ is not zero.  To resolve the disagreement, we must modify the bulk path integral so that adding a single $\tilde Z$ boundary gives a non-zero result which is independent of the number of boundaries already present. To motivate an appropriate modification, note that the  number of connected components in the bulk path integral is related to the dimension of the boundary Hilbert spaces, and so there may be a relation between the individual connected components and some basis for the CFT Hilbert space.
If so, to specify an eigenstate of $\hat{\tilde{Z}}$ we must specify which of these basis states are bosonic and which are fermionic. 
We can define projectors onto the bosonic and fermionic sectors by $1 \pm (-1)^F$, and we would like to insert such a projector into each of the connected components in the bulk path integral. 

This motivates the following proposal: we perform a bulk path integral with the bulk geometry restricted to have precisely $d$ connected components, and each connected component contains (in addition to any $Z$ or $\tilde Z$ boundaries) a boundary on which we sum over the spin structures, taking $NS + R$ if we want the corresponding state to be bosonic and $NS - R$ if we want the corresponding state to be fermionic.\footnote{Further inspiration for this picture came from the eigenbrane ideas of \cite{Saad:2019lba,Blommaert:2019wfy}.} 
We think of this boundary as a part of the specification of the dynamical bulk theory, rather than as part of the asymptotic boundary conditions we are free to choose. Accordingly, we treat connected components with no other boundaries apart from this one, the same genus and the same type of boundary (bosonic or fermionic), as indistinguishable, including a symmetry factor in the measure as in  \cite{Marolf:2020xie}.

\begin{figure}[t]
    \centering
    \includegraphics[scale=.165]{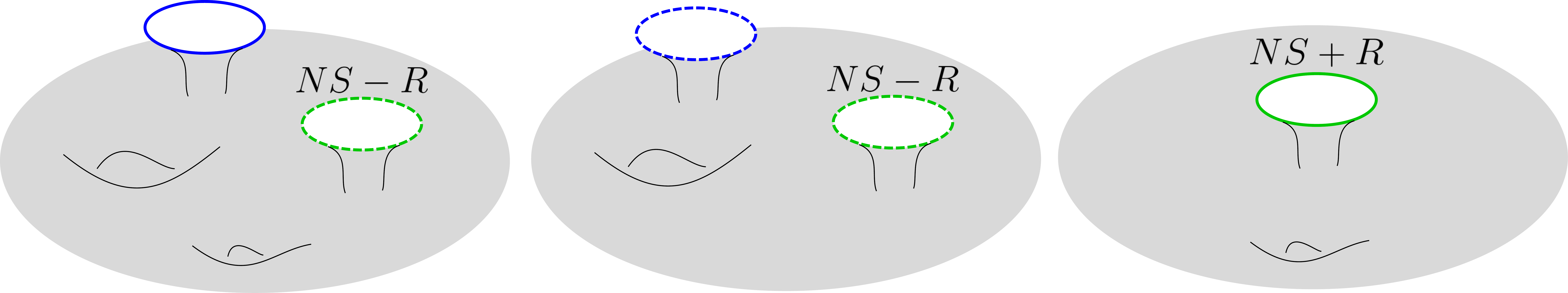}
    \caption{\small{A contribution to the expectation $\bra{Z=3,\tilde{Z}=-1} \hat{Z} \hat{\tilde{Z}} \ket{Z=3,\tilde{Z}=-1}$.  Since we wish to compute the expectation in an eigenstate, we must modify the usual rules for the gravitational path integral.  Since we want $Z=3$, the geometry must have exactly 3 connected components.  To select the $\tilde{Z}$ eigenstate, we choose specific sums over spin structures on the one additional boundary (green) on each component.  Choosing two $NS-R$ sums (dashed) and one $NS+R$ sum (solid) fixes $\tilde{Z} = -1$.  Finally, since the operator we want is $\hat{Z}\hat{\tilde{Z}}$, as usual we must fix the boundary conditions to have one NS boundary (solid blue) and one R boundary (dashed blue).}}
    \label{fig:single-theory}
\end{figure}

Including this additional contribution means we get a non-zero answer for any number of $\tilde Z$ boundaries in a connected component; if we have an even number we get a contribution from the NS spin structure on the additional boundary, and if we have an odd number we get a contribution from the R spin structure on the additional boundary. In the case with an $NS + R$ boundary,  adding a $\tilde Z$ boundary to a connected component gives a positive contribution, while, with a $NS-R$ boundary, we get a negative contribution.
With $d$ connected components of which $m$ have $NS - R$ boundaries, we get $d-m$ positive contributions and $m$ negative ones as we sum over different ways of adding the new $\tilde{Z}$ boundary.  So this corresponds to an eigenstate of $\tilde Z$ with eigenvalue $d-2m$ (figure \ref{fig:single-theory}). 

It is interesting to note that this means there are two potential bulk duals for a theory with equal numbers of bosonic and fermionic states:  we could consider the bulk path integral weighted by $(-1)^\zeta$ as described in section \ref{alt} and restrict to bulk geometries with $d$ connected components, or we could consider the path integral with trivial weighting, and restrict to $2d$ connected components, $d$ of which have a boundary summed over $NS+R$ spin structure and $d$ which have a boundary summed over $NS-R$ spin structure. These two bulk theories produce equivalent results. It would be interesting to understand the relation in more detail, and to see if there are other examples of different bulk constructions with the same boundary dual.

In addition to the above geometric construction of the dual of a single theory, we can also give an account of the eigenstates of $\hat{\tilde{Z}}$ which makes use of the ``spacetime D-branes" (SD-branes) of \cite{Marolf:2020xie}. Consider a theory where we allow the spacetime to have any number of additional SD-brane boundaries. We will allow for two types of SD boundaries, with NS or R spin structure, with complex parameters $g$, $\tilde g$ respectively. These SD-branes are not new objects; they are coherent states in the baby universe Hilbert space, 
\begin{equation}
    \ket{\text{SD-brane}_{g,\tilde{g}}} \equiv \ket{e^{gZ+\tilde{g}\tilde{Z}}} 
    = e^{g \hat{Z}+\tilde{g}\hat{\tilde{Z}}} \ket{\text{HH}},
\end{equation}
where $g,\tilde{g} \in \mathbb{C}$ are the SD-brane couplings. That is, calculating the path integral with SD-brane boundaries is equivalent to taking the expectation value in this coherent state. Considering the generating function
\begin{equation}
    \bra{\text{SD-brane}_{g,\tilde{g}}} e^{u\hat{Z}+\tilde{u}\hat{\tilde{Z}}} \ket{\text{SD-brane}_{g,\tilde{g}}} = \exp \left[ \lambda e^{u + 2 \text{Re}(g)} \cosh(\tilde{u} + 2\text{Re}(\tilde{g})) \right] ,
\end{equation}
where we used the previously derived relation $\ln \avg{e^{uZ+\tilde{u}\tilde{Z}}} = \lambda e^u \cosh \tilde{u}$.
To interpret this as an ensemble, we also need a new normalization factor
\begin{equation}
    \ln \mathfrak{Z}_{g,\tilde{g}} \equiv \ln \bra{\text{SD-brane}_{g,\tilde{g}}} 1 \ket{\text{SD-brane}_{g,\tilde{g}}} = \lambda e^{2\text{Re}(g)} \cosh 2\text{Re}(\tilde{g}) .
\end{equation}
Then we have an expansion of the generating function in the normalized SD-brane state
\begin{equation}
    \mathfrak{Z}^{-1}_{g,\tilde{g}} \bra{\text{SD-brane}_{g,\tilde{g}}} e^{u\hat{Z}+\tilde{u}\hat{\tilde{Z}}} \ket{\text{SD-brane}_{g,\tilde{g}}} = \sum_{d=0}^\infty p_d(\lambda') e^{ud} \sum_{m=0}^d b_{m|d}(\tilde{g}) e^{\tilde{u}(d-2m)},
\end{equation}
where $p_d(\lambda')$ is the Poisson distribution with shifted mean
\begin{equation}
    \lambda' \equiv \lambda e^{2\text{Re}(g)} \cosh 2\text{Re}(\tilde{g}) ,
\end{equation}
and $m$ for a given value of $d$ has a binomial distribution
\begin{equation}
b_{m|d}(\tilde{g}) \equiv (1+e^{-4\text{Re}(\tilde{g})})^{-d} e^{-4 m \text{Re}(\tilde{g})} \binom{d}{m} .
\end{equation}
We see that the $g$ dependence can be interpreted as a simple shift of the Poisson mean to $\lambda'$.
However, the $\tilde{g}$ dependence results in a more subtle modification, changing the relative probabilities of having fermionic or bosonic states. 
The collective distribution of $Z$ and $\tilde{Z}$ can be equivalently described as saying that the number of bosonic and fermionic states are independent random variables drawn from Poisson distributions with means $\lambda_b = \frac{1}{2} \lambda e^{2 \text{Re} (g) + 2 \text{Re} (\tilde g)}$ and $\lambda_f = \frac{1}{2} \lambda e^{2 \text{Re} (g) - 2 \text{Re} (\tilde g)}$ respectively. 

This is a rather interesting departure from the story in \cite{Marolf:2020xie}, where the total effect of the SD-brane was just the shift in the Poisson mean, which could also have been achieved by a suitable change in $\tilde{S}_0$.\footnote{Though we should remark that not all values of $\lambda \in \mathbb{R}_+$ are attainable by simply varying $\tilde{S}_0 \in \mathbb{R}$, so even in \cite{Marolf:2020xie} there is a small extension of the boundary ensemble due to SD-branes.  This is only true if we allow negative real SD-brane coupling $g < 0$.  If instead we require $g > 0$ then this subtlety is removed.}
Here, the SD-brane coherent state allows for a modification of the boundary ensemble which seems impossible to reach by only varying the constants that appear in the bulk gravitational action, as the bulk action does not distinguish bosonic and fermionic states.
If the general lesson to be learned about SD-brane boundary conditions for the bulk path integral is that they may lead to large modifications of the boundary ensemble, then perhaps a more clever application can actually cut down the number of parameters which define the boundary ensemble.

Since SD-brane boundaries correspond to coherent states in the baby universe Hilbert space, we can easily recover the eigenstates as appropriate combinations of them, by  Fourier transforming the SD-brane states for imaginary values of the couplings. We simply write
\begin{equation}
    \ket{FT[\text{SD-brane}]_{d,m}} \equiv \int_{-\pi}^\pi \frac{d\theta}{2\pi} \int_{-\pi}^\pi \frac{d\tilde{\theta}}{2\pi} e^{-id\theta} e^{-i(d-2m)\tilde{\theta}} \ket{\text{SD-brane}_{i\theta,i\tilde{\theta}}} ,
\end{equation}
with $d,m \in \mathbb{N}$.
Then by direct integration we find
\begin{equation}
    \ket{FT[\text{SD-brane}]_{d,m}} \propto \Big| \frac{\sin (\pi Z) \sin (\pi \tilde{Z})}{(Z-d)(\tilde{Z}-(d-2m))} \Big\rangle ,
\end{equation}
which are the eigenstates $\ket{Z=d,\tilde{Z}=d-2m}$ for $d$ and $m$ in the appropriate physical ranges.   As discussed above, these eigenstates are dual to a single boundary theory, rather than an ensemble of theories.

\section{JT gravity, spin structure, and EOW branes}
\label{jteow} 

Jackiw-Teitelboim (JT) gravity is a useful toy model of holography and quantum gravity which has led to many recent insights, including the possibility of holographic duality with an ensemble of theories (see, e.g.,  \cite{Almheiri:2014cka,Maldacena:2016upp,Almheiri:2019qdq,Penington:2019kki}, the references therein, and the review  \cite{Sarosi:2017ykf}). The sum over spin structures for JT gravity was extensively analysed in \cite{Stanford:2019vob}.  Here we follow our analysis of the topological model in previous sections to introduce EOW branes with spin structure, and to consider modifications of the bulk path integral to reproduce individual boundary theories.

We first consider adding EOW branes\footnote{See \cite{Kourkoulou:2017zaj,Penington:2019kki} for the construction and use of pure states by EOW branes in JT gravity, and \cite{Balasubramanian:2020hfs} for a higher dimensional discussion inspired by the ``inception" technique of \cite{Almheiri:2019hni}.} to the simple sum over spin structures, by combining the matrix model of  \cite{Stanford:2019vob} with the random state description of the EOW branes from \cite{Penington:2019kki}.  We consider JT gravity with the following boundary conditions: 
\begin{itemize}
    \item $p$ boundary segments of length $\tau_k$ connecting EOW branes with indices $i_k,j_k$, corresponding to $\langle \psi_{i_k} (\tau_k) | \psi_{j_k}(\tau_k) \rangle$, where $\ket{\psi(\tau)} \equiv e^{-\tau H/2} \ket{\psi}$,
    \item $q$ boundary segments of length $\tilde{\tau}_l$ connecting EOW branes with indices $\tilde{i}_l,\tilde{j}_l$, with an insertion of $(-1)^F$, corresponding to $\langle \psi_{\tilde{i}_l} (\tilde{\tau}_l) | (-1)^F | \psi_{\tilde{j}_l}(\tilde{\tau}_l) \rangle$,
    \item $r$ boundary circles of length $\beta_m$ with NS spin structure, corresponding to tr$(e^{-\beta_m H})$,
    \item and $s$ boundary circles of length $\tilde{\beta}_n$ with R spin structure, corresponding to \\ tr$((-1)^F e^{-\tilde{\beta}_n H})$.  
\end{itemize}

In the bulk path integral, we contract the $p+q$ boundary segments together in all possible ways with EOW branes. These pairings can be labelled by a permutation $\pi$ on $p+q$ variables, where the EOW brane starting in the $K$th boundary segment with index $j_K$ ends on the $\pi(K)$th boundary segment with index $i_{\pi(K)}$, where $K$ runs over the $p+q$ values labelled by $k,l$ in the above discussion. 
(We have appended the boundary R segment lengths to the boundary NS segment lengths, forming a long vector $\tau_K$, where $\tau_{p+l} \equiv \tilde{\tau}_l$.
We have done the same with the state indices, forming a long index vector $i_K$, where $i_{p+l} \equiv \tilde{i}_l$, and similarly for $j_K$, $j_{p+l}$, and $\tilde{j}_l$.)
The contribution vanishes when these indices don't match, so we have a sum over all permutations $\pi$  with a factor of $\prod_{K=1}^{p+q} \delta_{i_{\pi(K)}, j_K}$. For each choice of permutation, the boundary segments are sewn together to form a circular boundary for each cycle in the permutation $\pi$. When a cycle contains an even number of values lying in the range $K = p+1, \ldots p+q$, the boundary has Neveu-Schwarz spin structure, and when it contains an odd number, the boundary has Ramond spin structure. Each of these boundaries consists of an alternating sequence of EOW branes and asymptotic boundary segments, where the asymptotic boundary segments have length $\tau_K$, and the EOW branes follow geodesics in the bulk, with the length $l$ of the geodesic integrated over with a weight $e^{-\mu l}$ in the bulk path integral. 

\begin{figure}[t]
    \centering
    \includegraphics[scale=.15]{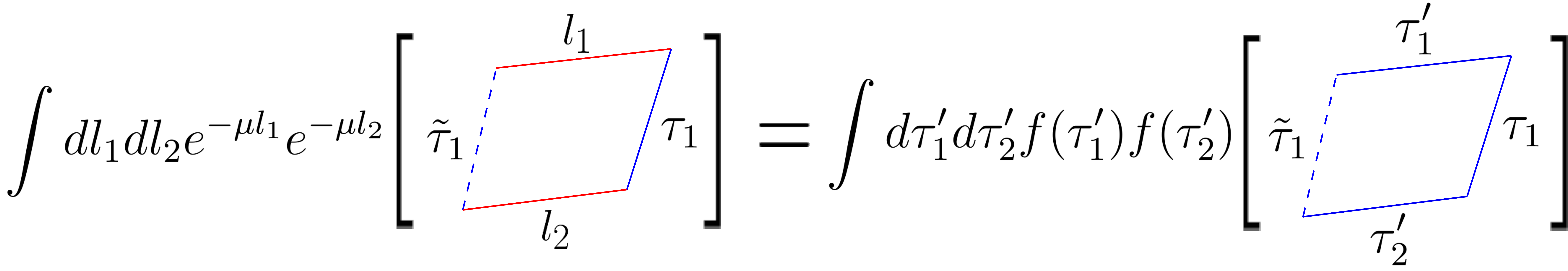}
    \caption{\small{An $N=2$ example of the identity in JT gravity relating the path integral with EOW branes and boundary segments forming a circular boundary to the path integral with only boundary segments forming a circular boundary.  The red EOW brane segments which have tension $\mu$ and lengths $l_i$ turn into NS boundary segments of length $\tau_i'$.  For the equality to hold, we must integrate over the EOW brane lengths and transformed NS boundary segment lengths with the weights shown, given in \eqref{eq:brane-weights}. This transformation leaves the spin structure unchanged; here it is Ramond, since we have an odd number of dashed segments.}}
    \label{fig:eow-replace}
\end{figure}

In Appendix D in \cite{Penington:2019kki}, Penington et al use previous results relating the bulk path integral with a boundary segment of length $\tau'$ to a bulk path integral with a boundary along a bulk geodesic to show that the boundary condition with (say) $N$ alternating EOW branes and boundary segments, where the boundary segments have length $\tau_i$, $i= 1, \ldots N$, is equivalent to taking the boundary condition with a circular boundary of length $\sum_i (\tau_i + \tau'_i)$, and integrating over all the $\tau'_i$ with weights 
\begin{equation} 
f(\tau'_i) = IL \left[ 2^{1-2\mu} \Bigg| \Gamma \left( \mu-\frac{1}{2} + i \sqrt{2E_a}  \right) \Bigg|^2 \right](\tau'_i),    
\label{eq:brane-weights}
\end{equation}
where IL denotes the inverse Laplace transform.
The spin structure is unaffected by this transformation, so if we had an even number of the $(-1)^F$ boundary segments in the EOW brane boundary, we will have an NS spin structure on the circular boundary, and if we have an odd number, we will have a R spin structure (figure \ref{fig:eow-replace}). 

Thus, we can reduce the evaluation of the bulk path integral for these boundary conditions to a sum of terms each of which involves some number of circular boundaries of given length with NS and R spin structures; precisely the case that was analysed in \cite{Stanford:2019vob}. 

The calculation there was organised by considering the genus $g$ connected contribution to the bulk path integral with these boundary conditions, with $n$ total boundaries of which $n_R$ have Ramond boundary conditions, which they call $Z_{g,n,n_R}$ (after extracting a factor of $e^{-(2g+n-2)S_0}$, and rescaling so that $Z_{0,1,0}=1$). They show that summing over all spin structures gives
\begin{equation} \label{jt1} 
Z_{g,n,n_R}= 2^{2g+n-2} (1+(-1)^{n_R}) Z^{GUE}_{g,n}     
\end{equation}
and summing over spin structures with $(-1)^\zeta$ gives 
\begin{equation}
    Z_{g,n,n_R} =
    \begin{cases}
        2^{2g+n-1} Z^{GUE}_{g,n}, & n_R=0 \\
        0, & \text{else},
    \end{cases}
\end{equation}
where $Z^{GUE}_{g,n}$ denotes the result without a sum over spin structures, which we know from \cite{Saad:2019lba} is equal to the result in the GUE matrix ensemble.\footnote{We follow the notation of \cite{Stanford:2019vob}, where GUE refers to a generalized class of unitary-invariant matrix ensembles which need not be precisely Gaussian.  Indeed, the JT ensemble has a leading double-scaled density of states which looks like $\sinh \sqrt{E}$, which is quite far from the Gaussian result $\sqrt{E}$.}

Let us now see how we reproduce this result from an ensemble perspective. For the case where we simply sum over spin structures, the situation is straightforward. The sum over spin structures in JT gravity is dual to a random matrix ensemble \cite{Stanford:2019vob}
\begin{equation} \label{Hens}
H = \left( \begin{array}{cc} GUE_1 & 0 \\ 0 & GUE_2 \end{array} \right),
\end{equation}
where there are independent random matrices for the Hamiltonian acting on the bosonic states (the upper block) and the fermionic states (the lower block). In \cite{Penington:2019kki}, the EOW brane states are taken to be random superpositions of the energy eigenstates. The natural generalization is 
\begin{equation} \label{rst}
|\psi_i(\tau) \rangle = \left( \sum_b + \sum_f \right) 2^{\frac{1}{2} - \mu} \Gamma \left( \mu - \frac{1}{2} + i \sqrt{2E_a} \right) e^{-\tau E_a} C_{i,a} | E_a \rangle, 
\end{equation}
where $\sum_b + \sum_f$ indicates that we sum the index $a$ over both the bosonic and fermionic states, that is over all eigenstates of $H$ above, and $C_{i,a}$ are i.i.d. complex Gaussian random variables. 
Intuitively, this formula follows from \eqref{eq:brane-weights} and the fact that the inverse Laplace transform of the partition function yields a density of states.
Then we also have
\begin{equation} \label{rstf}
(-1)^F |\psi_{\tilde{i}}(\tilde{\tau}) \rangle = \left( \sum_b - \sum_f \right) 2^{\frac{1}{2} - \mu} \Gamma \left( \mu - \frac{1}{2} + i \sqrt{2E_a} \right) e^{-\tilde{\tau} E_a} C_{\tilde{i},a} | E_a \rangle. 
\end{equation}
Now we consider the average over $H,C$ of the observable with $p$ normal boundary segments, $q$ $(-1)^F$ boundary segments, $r$ normal circles and $s$ $(-1)^F$ circles (where we now use $\mathbb{E}$ for the boundary ensemble expectation instead of $\avg{\cdot}$, since the bra-ket now labels the inner product in the boundary Hilbert space $\mathcal{H}$, i.e. it is playing the role of the parenthesis $(\psi_i,\psi_j)$ in the topological model)
\begin{equation} \label{obs}
\mathbb{E}_{H,C} \left( \prod_{k=1}^p \langle \psi_{i_k} (\tau_k) | \psi_{j_k}(\tau_k) \rangle \prod_{l=1}^q \langle \psi_{\tilde{i}_l} (\tilde{\tau}_l) | (-1)^F | \psi_{\tilde{j}_l}(\tilde{\tau}_l) \rangle \prod_{m=1}^r \mathrm{tr}(e^{-\beta_m H}) \prod_{n=1}^s \mathrm{tr} ((-1)^F e^{-\tilde{\beta}_n H}) \right). 
\end{equation}
The average over the $C_{i,a}$ generates contractions between $|\psi_j\rangle$ and $\langle \psi_i |$, producing a sum over permutations of the $p+q$ elements just as in the bulk analysis, so \eqref{obs} equals
\begin{equation} 
\sum_\pi \left\{ \prod_{K=1}^{p+q} \delta_{i_{\pi(K)}, j_K} \mathbb{E}_{H} \left[ \prod_{\gamma \in c(\pi)} \mathrm{tr} \left( \rho_\gamma \right)  \prod_{m=1}^r \mathrm{tr}(e^{-\beta_m H}) \prod_{n=1}^s \mathrm{tr} ((-1)^F e^{-\tilde{\beta}_n H}) \right] \right\},
\end{equation}
where
\begin{equation} 
\rho_\gamma = \sum_a (-1)^{n_q F}  e^{-\sum_{L \in \gamma} \tau_L H}  |\Gamma(\mu- \frac{1}{2} + i \sqrt{2 E_a})|^2 |E_a \rangle \langle E_a |.
 \end{equation}
There is a trace over the Hilbert space for each cycle in the permutation, coming from the remaining sum over the indices in \eqref{rst} and \eqref{rstf} after contracting all the $C_{i,a}$. This comes with a minus sign for the fermionic states if the cycle involved an odd number of factors of $(-1)^F$, which is accounted for by the factor of $(-1)^{n_q F}$, where $n_q$ is the number of values from $K=p+1, \ldots p+q$ appearing in the cycle. If this is an even number we get the normal trace and if it's an odd number we get the trace with $(-1)^F$. 

Using the inverse Laplace transform, this is equal to 
\begin{equation} \label{mres} 
\begin{split}
\sum_\pi \int & \prod_K d\tau'_K f(\tau'_K) \delta_{i_{\pi(K)}, j_K} \times \\ & \mathbb{E}_{H} \left[ \prod_{\gamma \in c(\pi)} \mathrm{tr} \left(  (-1)^{n_q F}  e^{-\sum_{L \in \gamma} (\tau_L+\tau'_L) H}  \right)  \prod_{m=1}^r \mathrm{tr}(e^{-\beta_m H}) \prod_{n=1}^s \mathrm{tr} ((-1)^F e^{-\tilde{\beta}_n H}) \right]. 
\end{split}
\end{equation}
This is the expectation value in the matrix ensemble of some number of circular boundaries, with a permutation factor and an integral over the auxiliary variables $\tau'_K$ which exactly matches what we argued above will appear in the bulk path integral. The average over the matrix ensemble defined by 
\eqref{Hens} exactly reproduces the result \eqref{jt1} for the values of the bulk path integral with circular boundary conditions \cite{Stanford:2019vob}. Thus, this ensemble of random matrices and random states precisely matches the bulk JT gravity path integral results with a sum over spin structures for the EOW brane boundary conditions described at the start of this section. 

Now let's consider the sum over spin structures with an insertion of $(-1)^\zeta$. In \cite{Stanford:2019vob} this was related to a matrix ensemble with no symmetry, so $H = GUE$. But as in section \ref{alt}, once we introduce the EOW branes, we need the boundary Hilbert space to have some additional structure to account for the difference between $\langle \psi_{i_k} (\tau_k) | \psi_{j_k}(\tau_k) \rangle$ and $\langle \psi_{\tilde{i}_l} (\tilde{\tau}_k) | (-1)^F | \psi_{\tilde{j}_l}(\tilde{\tau}_k) \rangle$ boundary segments. A natural proposal extending the idea of the previous section is to keep a single matrix ensemble, but double the Hilbert space, so 
\begin{equation} \label{Hs} 
H = \left( \begin{array}{cc} GUE & 0 \\ 0 & GUE \end{array} \right),
\end{equation}
where we interpret the upper block as bosonic states and the lower block as fermionic states, as in our discussion of the usual sum over spin structures. The result is then the same as \eqref{mres}, but averaging over the matrix model will make the answer vanish unless $s=0$, and we restrict to permutations with even numbers of $(-1)^F$ elements in each cycle, so the matrix model answer is equal to 
\begin{equation} 
\sum_{\pi'} \int \prod_K d\tau'_K f(\tau'_K) \delta_{i_{\pi'(K)}, j_K} \mathbb{E}_{H} \left[ \prod_{\gamma \in c(\pi')} \mathrm{tr} \left(   e^{-\sum_{L \in \gamma} (\tau_L+\tau'_L) H}  \right)  \prod_{m=1}^r \mathrm{tr}(e^{-\beta_m H}) \right],
\end{equation}
where $\sum_{\pi'}$ denotes the restricted sum over permutations. This makes the answer zero in the appropriate cases, and almost gives the expected answer for $Z_{NS}(\beta)$ and the traces with an even number of boundaries. However, as in section \ref{alt}, there is an extra factor of 2 from the doubling of the Hilbert space. We can deal with this by hand by identifying the bulk objects with rescaled versions of the matrix model observables, so $Z_{NS}(\beta) = \frac{1}{\sqrt{2}} \mathrm{tr} e^{-\beta H}$, rather than the identification in \cite{Stanford:2019vob} of $Z_{NS}(\beta) = \sqrt{2} \mathrm{tr} e^{-\beta H}$. This factor seems a little ad hoc, and it would be nice to have a deeper understanding of it. 

The proposal in \cite{Stanford:2019vob} was motivated by considering the SYK model. For even $N$ the SYK Hilbert space has a $(-1)^F$ symmetry, but for odd $N$ $(-1)^F$ interchanges two irreducible representations of the Clifford algebra, so if we take the Hilbert space to be one of these two representations the symmetry is broken at the quantum level. Our proposal could be understood as corresponding to keeping both representations, so that the boundary Hilbert space is the direct sum. The bosonic and fermionic states are symmetric and antisymmetric combinations of the states exchanged by $(-1)^F$.

This model has a degeneracy between the bosonic and fermionic states, since the Hamiltonians in the two sectors are identified. Thus, each instance is a supersymmetric quantum mechanics. However, this is not the same as the supersymmetric matrix model, as we are taking the Hamiltonian rather than the supercharge as the random variable. For each instance we can construct a supercharge corresponding to the Hamiltonian \eqref{Hs}, but averaging over the GUE in \eqref{Hs} is not the same as averaging over a GUE ensemble for the supercharge. This is why the bulk description, corresponding to an ensemble averaged theory, is simply JT gravity and not super JT.

Finally, we consider how we can modify the bulk path integral to produce a dual of a specific boundary theory. For JT gravity, it was already suggested at the level of the purely bosonic theory without spin structures that fixing the spectrum of the boundary Hamiltonian would involve introducing additional boundaries in the bulk path integral; indeed this was part of the inspiration for our proposal in section \ref{single}. In particular, in \cite{Blommaert:2019wfy}, it was argued that introducing ``eigenbrane" boundaries (analogous to the FZZT branes of \cite{Maldacena:2004sn}) in the bulk could be associated with having some fixed eigenvalues. 
This perspective was further explored in \cite{Blommaert:2020seb}.
In the model with a sum over spin structures, we want to specify whether the fixed eigenvalue associated with a given eigenbrane boundary is bosonic or fermionic. It seems natural to do so by summing over spin structures on the eigenbrane boundary as in section \ref{single}, with $NS+R$ boundary spin structure giving a bosonic eigenvalue and $NS-R$ spin structure giving a fermionic eigenvalue. 

From the boundary ensemble average point of view, it is clear what sort of operators we must introduce.
For the standard sum over spin structures, we have a boundary ensemble with  two separate random matrices, and the partition function is schematically
\begin{equation}
    \mathcal{Z}_{\text{standard}} = \int dH_+ \int dH_-\ e^{-LV(H_+)-LV(H_-)} ,
\end{equation}
where $H_+$ is the bosonic sector Hamiltonian, $H_-$ is the fermionic sector Hamiltonian, and $V$ is the potential.
In such a situation, we can introduce bosonic and fermionic eigenbrane operators
\begin{equation}
    \psi_\pm(E) \equiv e^{-\frac{L V(E)}{2}} \det (E-H_\pm) .
\end{equation}
Then, incorporating a bosonic or fermionic boundary with fixed energy $\lambda$ corresponds to a matrix ensemble where the partition function involves an expectation over $\psi^2_\pm(\lambda)$, which fixes a single eigenvalue in the appropriate sector as in \cite{Blommaert:2019wfy}.
By contrast, in the alternate sum over spin structures, there is effectively only one random matrix to average over, and we have $\psi_+(E) = \psi_-(E) \equiv \psi(E)$, which reduces identically to the situation considered in \cite{Blommaert:2019wfy}.
Of course, in both of these situations, the equivalence between the boundary ensemble picture and the bulk sum over surfaces (with extra fixed energy boundaries) picture is still guaranteed by the equivalence of JT gravity's genus expansion and the matrix model recursion relations as proven in \cite{Saad:2019lba}.

\section{Conclusions}
\label{concl}

In this paper, we  extended the simple topological model of \cite{Marolf:2020xie} to include a sum over spin structures. The gravitational path integral then has a dual interpretation as an ensemble average over theories labelled by the dimension of a boundary Hilbert space, with an additional decomposition into bosonic and fermionic states. If all spin structures are equally weighted, the states in the dual are chosen to be bosonic or fermionic at random in the different theories in the ensemble. This  freedom effectively enlarges the baby universe Hilbert space. An alternative sum over spin structures weighted by $(-1)^\zeta$ (where $\zeta$ counts zero modes mod 2 of the Dirac equation) leads to a dual ensemble in which the number of bosonic and fermionic states is fixed to be equal, and the corresponding baby universe Hilbert space is the same as in  \cite{Marolf:2020xie}  without a sum over spin structures. We discussed the bulk dual of individual boundary theories in the ensemble, and argued that this could be obtained by restricting the bulk path integral to geometries with a fixed number of connected components, with an additional boundary in each connected component on which we sum  over spin structures.

We can add EOW branes as a form of topological matter: the boundary conditions then include  boundary segments with standard and twisted boundary conditions, which carry state labels associated to the EOW branes that terminate them.  With this addition, the gravitational path integral has a dual ensemble description in which the boundary states are random linear combinations of the bosonic and fermionic elements of the boundary Hilbert space.  We also discussed the calculation with EOW branes for JT gravity. The theory with the sum over spin structures weighted by $(-1)^\zeta$ can be interpreted as dual to a matrix model with a degeneracy between fermionic and bosonic states; the individual instances of the model are supersymmetric, but the average is over choices of the Hamiltonian (from a GUE ensemble) rather than the supercharge, so this is not a supersymmetric matrix model as usually understood. 

Extending the model by adding additional structures in the bulk generically increases the size of the baby universe Hilbert space. The number of null states  also increases, as there are restrictions on the values of the boundary observables such as the Witten index $\tilde Z$, which imply additional null states. But boundary observables typically do have a range of possible values, which imply new physical baby universe states. In some definitions of the gravitational path integral, e.g., if the  sum over spin structures is weighted by $(-1)^\zeta$,  the new observables have a unique value, where $\tilde Z =0$.  In this case, the baby universe Hilbert space is not reduced compared to the case without summing over spin structures; it is just not enlarged. From the boundary perspective, it seems likely that this behaviour is generic. If an  extension adds new structures in the boundary Hilbert space, we have a larger space of possible boundary models consistent with the structure, and the dual gravitational  path integral in the bulk is dual to a sum over all models with this structure. 

To produce a bulk path integral which is dual to a unique boundary theory, or perhaps a more restricted class of models, we must implement an analogous restriction from the bulk perspective. It is far from clear how to achieve this in general. In our discussion of the simple topological model and JT gravity, it was possible to achieve this by adding additional boundary components in the bulk path integral with particular weightings.   It would be interesting to extend this discussion
to obtain a path integral dual of the calculation in a specific baby universe state (or equivalently a unique boundary theory) for the theory with EOW branes. This is particularly interesting as it provides the simplest example where the eigenstates of the boundary creation operators are not normalisable states in the baby universe Hilbert space - they are only delta function normalisable. 

One attraction of adding spin structures to the simple toy model is that we get a topological obstruction to the existence of bulk geometries for certain boundary conditions, which then implies that the bulk path integral must have an ensemble interpretation. Looking for such topological obstructions is a good way to explore whether the bulk gravity calculation has an ensemble interpretation in other, more complicated cases, where we cannot control the full gravitational path integral. 

It has been conjectured \cite{McNamara:2020uza} that the baby universe Hilbert space is one-dimensional in consistent theories of quantum gravity, such as those obtained from string theory, which would imply the absence of any such topological obstruction. 
In a discussion of compactifications, the same authors conjectured earlier that any compact $d$-manifold is the boundary of some $(d+1)$-dimensional dynamical process in a consistent theory of quantum gravity \cite{McNamara:2019rup}. Similar topological issues arise in discussions of anomalies (see e.g. \cite{Garcia-Etxebarria:2018ajm})  and in a recent study of bubbles of nothing \cite{GarciaEtxebarria:2020xsr}. 

Mathematically, the relevant structure is a cobordism group, specifically the $d$-dimensional spin cobordism groups $\Omega_d^{\text{Spin}}$, where $d$ is the dimension of the boundary and disjoint union is the group operation \cite{Milnor:1963}. 
Elements of $\Omega_d^{\text{Spin}}$ are equivalence classes of $d$-manifolds $M$ with spin structure $\sigma$ such that two manifolds $(M_1,\sigma_1)$ and $(M_2,\sigma_2)$ are equivalent if there is a $(d+1)$-manifold $\mathcal{M}$ with spin structure $\Sigma$ and $\partial \mathcal{M} = M_1 \sqcup M_2$ and $\Sigma$ induces $\sigma_1$ on the $M_1$ boundary and $\sigma_2$ on the $M_2$ boundary. 
For the $d=1$ case we considered here, $\Omega_1^{\text{Spin}} = \mathbb{Z}_2$, and $S^1$ with Ramond structure (periodic boundary conditions for fermions) is the nontrivial element; thus, a two-dimensional spin manifold always has an even number of R boundaries, as claimed previously. 

There is a direct analogue for $d=2$, where $\Omega_2^{\text{Spin}} = \mathbb{Z}_2$  and the torus $S^1 \times S^1$ with R structure on both circles is the nontrivial element. Thus, if we considered an AdS$_3/$CFT$_2$ correspondence with the CFT living on this torus, it looks like there is a similar obstruction. However, top-down AdS$_3/$CFT$_2$ models usually have some internal compact space. If we consider a model with AdS$_3 \times S^3$ asymptotics, the boundary at some cutoff surface is $T^2 \times S^3$. The relevant spin cobordism group is then actually $\Omega_5^{\text{Spin}}$, which is trivial. This is easy to understand; one can construct a spin manifold by filling in the $S^3$. Similarly, in $d=4$, there is a non-trivial  $\Omega_4^{\text{Spin}} = \mathbb{Z}$ which is generated by the $K3$ surface. But in the duality with $\mathcal N=4$ SYM, we need to consider $K3 \times S^5$, and there are ten-dimensional spin manifolds where we fill in the $S^5$. This relaxation of the topological restriction can also arise in more top-down constructions of two-dimensional models. In the topological theory we considered and in JT gravity, the bulk is actually two-dimensional, but in a string theory construction, we often obtain AdS$_2$ as the near-horizon region of some black hole, and the full geometry has an AdS$_2 \times S^d$ factor; there is then no obstruction to having a spin manifold with an $S^1 \times S^d$ boundary for periodic spin structure on the $S^1$. 

Thus, we see that familiar top-down constructions do not have topological obstructions similar to the one in our simple two-dimensional model. This is far from a systematic exploration of the possibilities, however, and it remains an interesting direction for future work to see if candidate topological obstructions can be identified in other cases.

The authors of  \cite{Balasubramanian:2006sg} gave an
interesting  example of a statistical ensemble of theories appearing in the description of a fully dynamical universe with a spacetime boundary.  They described a universe with a sort of Big Bang seeded by an unstable brane placed at the beginning of time.  The brane decays and populates the universe with quanta. The entire process can be described by open string worldsheet computations, and, equivalently, in a grand canonical ensemble of $SU(N)$ matrix models of different ranks $N$. Later points in time are related to larger chemical potentials and thus to dominance by the dynamics of larger matrices, consistent with the idea that large matrix models tend to have dual gravitational descriptions.   In this way, time emerges from a statistical ensemble on a Euclidean surface, and the size of space is related to the number of degrees of freedom, with early times (where the brane tension should shrink space) related to the contribution of  small matrices, and late times (where space is large) being related to the contribution of  large matrices. There are some analogies with the topological model discussed in this paper.  Here, too, we have an ensemble of finite dimensional quantum theories which collectively reproduce computations in bulk gravity.  We have also seen, following \cite{Marolf:2020xie}, that the dimension of the boundary Hilbert space is related to the number of connected components in the bulk geometry.  In a topological theory there is no metric, and so, in some sense, the number of connected components is a proxy for the size of the universe.  So, like in  \cite{Balasubramanian:2006sg}, pieces of the boundary ensemble with Hilbert spaces of different dimension describe components of the universe of different size.
It would be interesting to understand if there is a deeper connection between these two pictures.

\section*{Acknowledgements}

We thank Matt DeCross, Cathy Li, and G\'{a}bor S\'{a}rosi for useful discussions.
SFR is supported in part by STFC through grant ST/P000371/1. 
VB was supported in part by the DOE grants FG02-05ER-41367 and QuantISED grant DE-SC0020360.
VB, TU, and AK were supported in part by the Simons
Foundation through the It From Qubit Collaboration (Grant No. 38559).
VB and AK were supported in part by DOE grant DE-SC0013528. TU was supported by JSPS Grant-in-Aid for Young Scientists  19K14716.

\bibliographystyle{JHEP}
\bibliography{euclidean}

\providecommand{\href}[2]{#2}\begingroup\raggedright\begin{thebibliography}{10}

\bibitem{Lavrelashvili:1987jg}
G.~V. Lavrelashvili, V.~Rubakov and P.~Tinyakov, \emph{{Disruption of Quantum
  Coherence upon a Change in Spatial Topology in Quantum Gravity}}, {\emph{JETP
  Lett.} {\bf 46} (1987) 167--169}.

\bibitem{Hawking:1987mz}
S.~Hawking, \emph{{Quantum Coherence Down the Wormhole}},
  \href{http://dx.doi.org/10.1016/0370-2693(87)90028-1}{\emph{Phys. Lett. B}
  {\bf 195} (1987) 337}.

\bibitem{Giddings:1987cg}
S.~B. Giddings and A.~Strominger, \emph{{Axion Induced Topology Change in
  Quantum Gravity and String Theory}},
  \href{http://dx.doi.org/10.1016/0550-3213(88)90446-4}{\emph{Nucl. Phys. B}
  {\bf 306} (1988) 890--907}.

\bibitem{Hawking:1988ae}
S.~Hawking, \emph{{Wormholes in Space-Time}},
  \href{http://dx.doi.org/10.1103/PhysRevD.37.904}{\emph{Phys. Rev. D} {\bf 37}
  (1988) 904--910}.

\bibitem{Coleman:1988cy}
S.~R. Coleman, \emph{{Black Holes as Red Herrings: Topological Fluctuations and
  the Loss of Quantum Coherence}},
  \href{http://dx.doi.org/10.1016/0550-3213(88)90110-1}{\emph{Nucl. Phys.} {\bf
  B307} (1988) 867--882}.

\bibitem{Giddings:1988cx}
S.~B. Giddings and A.~Strominger, \emph{{Loss of Incoherence and Determination
  of Coupling Constants in Quantum Gravity}},
  \href{http://dx.doi.org/10.1016/0550-3213(88)90109-5}{\emph{Nucl. Phys.} {\bf
  B307} (1988) 854--866}.

\bibitem{Giddings:1988wv}
S.~B. Giddings and A.~Strominger, \emph{{Baby Universes, Third Quantization and
  the Cosmological Constant}},
  \href{http://dx.doi.org/10.1016/0550-3213(89)90353-2}{\emph{Nucl. Phys.} {\bf
  B321} (1989) 481--508}.

\bibitem{Maldacena:2004rf}
J.~M. Maldacena and L.~Maoz, \emph{{Wormholes in AdS}},
  \href{http://dx.doi.org/10.1088/1126-6708/2004/02/053}{\emph{JHEP} {\bf 02}
  (2004) 053}, [\href{https://arxiv.org/abs/hep-th/0401024}{{\tt
  hep-th/0401024}}].

\bibitem{Hawking:1974sw}
S.~Hawking, \emph{{Particle Creation by Black Holes}},
  \href{http://dx.doi.org/10.1007/BF02345020}{\emph{Commun. Math. Phys.} {\bf
  43} (1975) 199--220}.

\bibitem{Almheiri:2019qdq}
A.~Almheiri, T.~Hartman, J.~Maldacena, E.~Shaghoulian and A.~Tajdini,
  \emph{{Replica Wormholes and the Entropy of Hawking Radiation}},
  \href{http://dx.doi.org/10.1007/JHEP05(2020)013}{\emph{JHEP} {\bf 05} (2020)
  013}, [\href{https://arxiv.org/abs/1911.12333}{{\tt 1911.12333}}].

\bibitem{Almheiri:2019hni}
A.~Almheiri, R.~Mahajan, J.~Maldacena and Y.~Zhao, \emph{{The Page curve of
  Hawking radiation from semiclassical geometry}},
  \href{http://dx.doi.org/10.1007/JHEP03(2020)149}{\emph{JHEP} {\bf 03} (2020)
  149}, [\href{https://arxiv.org/abs/1908.10996}{{\tt 1908.10996}}].

\bibitem{Penington:2019kki}
G.~Penington, S.~H. Shenker, D.~Stanford and Z.~Yang, \emph{{Replica wormholes
  and the black hole interior}},  \href{https://arxiv.org/abs/1911.11977}{{\tt
  1911.11977}}.

\bibitem{Almheiri:2019psy}
A.~Almheiri, R.~Mahajan and J.~E. Santos, \emph{{Entanglement islands in higher
  dimensions}},  \href{https://arxiv.org/abs/1911.09666}{{\tt 1911.09666}}.

\bibitem{Balasubramanian:2020hfs}
V.~Balasubramanian, A.~Kar, O.~Parrikar, G.~Sárosi and T.~Ugajin,
  \emph{{Geometric secret sharing in a model of Hawking radiation}},
  \href{https://arxiv.org/abs/2003.05448}{{\tt 2003.05448}}.

\bibitem{Almheiri:2020cfm}
A.~Almheiri, T.~Hartman, J.~Maldacena, E.~Shaghoulian and A.~Tajdini,
  \emph{{The entropy of Hawking radiation}},
  \href{https://arxiv.org/abs/2006.06872}{{\tt 2006.06872}}.

\bibitem{Polchinski:1994zs}
J.~Polchinski and A.~Strominger, \emph{{A Possible resolution of the black hole
  information puzzle}},
  \href{http://dx.doi.org/10.1103/PhysRevD.50.7403}{\emph{Phys. Rev. D} {\bf
  50} (1994) 7403--7409}, [\href{https://arxiv.org/abs/hep-th/9407008}{{\tt
  hep-th/9407008}}].

\bibitem{Coleman:1988tj}
S.~R. Coleman, \emph{{Why There Is Nothing Rather Than Something: A Theory of
  the Cosmological Constant}},
  \href{http://dx.doi.org/10.1016/0550-3213(88)90097-1}{\emph{Nucl. Phys. B}
  {\bf 310} (1988) 643--668}.

\bibitem{Preskill:1988na}
J.~Preskill, \emph{{Wormholes in Space-time and the Constants of Nature}},
  \href{http://dx.doi.org/10.1016/0550-3213(89)90592-0}{\emph{Nucl. Phys. B}
  {\bf 323} (1989) 141--186}.

\bibitem{Klebanov:1988eh}
I.~R. Klebanov, L.~Susskind and T.~Banks, \emph{{Wormholes and the Cosmological
  Constant}}, \href{http://dx.doi.org/10.1016/0550-3213(89)90538-5}{\emph{Nucl.
  Phys. B} {\bf 317} (1989) 665--692}.

\bibitem{Saad:2019lba}
P.~Saad, S.~H. Shenker and D.~Stanford, \emph{{JT gravity as a matrix
  integral}},  \href{https://arxiv.org/abs/1903.11115}{{\tt 1903.11115}}.

\bibitem{Maloney:2020nni}
A.~Maloney and E.~Witten, \emph{{Averaging Over Narain Moduli Space}},
  \href{https://arxiv.org/abs/2006.04855}{{\tt 2006.04855}}.

\bibitem{Afkhami-Jeddi:2020ezh}
N.~Afkhami-Jeddi, H.~Cohn, T.~Hartman and A.~Tajdini, \emph{{Free partition
  functions and an averaged holographic duality}},
  \href{https://arxiv.org/abs/2006.04839}{{\tt 2006.04839}}.

\bibitem{Belin:2020hea}
A.~Belin and J.~de~Boer, \emph{{Random Statistics of OPE Coefficients and
  Euclidean Wormholes}},  \href{https://arxiv.org/abs/2006.05499}{{\tt
  2006.05499}}.

\bibitem{Cotler:2020ugk}
J.~Cotler and K.~Jensen, \emph{{AdS$_3$ gravity and random CFT}},
  \href{https://arxiv.org/abs/2006.08648}{{\tt 2006.08648}}.

\bibitem{Bousso:2020kmy}
R.~Bousso and E.~Wildenhain, \emph{{Gravity/Ensemble Duality}},
  \href{https://arxiv.org/abs/2006.16289}{{\tt 2006.16289}}.

\bibitem{Marolf:2020xie}
D.~Marolf and H.~Maxfield, \emph{{Transcending the ensemble: baby universes,
  spacetime wormholes, and the order and disorder of black hole information}},
  \href{https://arxiv.org/abs/2002.08950}{{\tt 2002.08950}}.

\bibitem{Hartle:1983ai}
J.~Hartle and S.~Hawking, \emph{{Wave Function of the Universe}},
  \href{http://dx.doi.org/10.1103/PhysRevD.28.2960}{\emph{Adv. Ser. Astrophys.
  Cosmol.} {\bf 3} (1987) 174--189}.

\bibitem{Anous:2020lka}
T.~Anous, J.~Kruthoff and R.~Mahajan, \emph{{Density matrices in quantum
  gravity}},  \href{https://arxiv.org/abs/2006.17000}{{\tt 2006.17000}}.

\bibitem{Stanford:2019vob}
D.~Stanford and E.~Witten, \emph{{JT Gravity and the Ensembles of Random Matrix
  Theory}},  \href{https://arxiv.org/abs/1907.03363}{{\tt 1907.03363}}.

\bibitem{Blommaert:2019wfy}
A.~Blommaert, T.~G. Mertens and H.~Verschelde, \emph{{Eigenbranes in
  Jackiw-Teitelboim gravity}},  \href{https://arxiv.org/abs/1911.11603}{{\tt
  1911.11603}}.

\bibitem{McNamara:2020uza}
J.~McNamara and C.~Vafa, \emph{{Baby Universes, Holography, and the
  Swampland}},  \href{https://arxiv.org/abs/2004.06738}{{\tt 2004.06738}}.

\bibitem{Gesteau:2020wrk}
E.~Gesteau and M.~J. Kang, \emph{{Holographic baby universes: an observable
  story}},  \href{https://arxiv.org/abs/2006.14620}{{\tt 2006.14620}}.

\bibitem{Kourkoulou:2017zaj}
I.~Kourkoulou and J.~Maldacena, \emph{{Pure states in the SYK model and
  nearly-$AdS_2$ gravity}},  \href{https://arxiv.org/abs/1707.02325}{{\tt
  1707.02325}}.

\bibitem{Dijkgraaf:2018vnm}
R.~Dijkgraaf and E.~Witten, \emph{{Developments in Topological Gravity}},
  \href{http://dx.doi.org/10.1142/S0217751X18300296}{\emph{Int. J. Mod. Phys.
  A} {\bf 33} (2018) 1830029}, [\href{https://arxiv.org/abs/1804.03275}{{\tt
  1804.03275}}].

\bibitem{Giddings:2020yes}
S.~B. Giddings and G.~J. Turiaci, \emph{{Wormhole calculus, replicas, and
  entropies}},  \href{https://arxiv.org/abs/2004.02900}{{\tt 2004.02900}}.

\bibitem{Almheiri:2014cka}
A.~Almheiri and J.~Polchinski, \emph{{Models of AdS$_{2}$ backreaction and
  holography}}, \href{http://dx.doi.org/10.1007/JHEP11(2015)014}{\emph{JHEP}
  {\bf 11} (2015) 014}, [\href{https://arxiv.org/abs/1402.6334}{{\tt
  1402.6334}}].

\bibitem{Maldacena:2016upp}
J.~Maldacena, D.~Stanford and Z.~Yang, \emph{{Conformal symmetry and its
  breaking in two dimensional Nearly Anti-de-Sitter space}},
  \href{http://dx.doi.org/10.1093/ptep/ptw124}{\emph{PTEP} {\bf 2016} (2016)
  12C104}, [\href{https://arxiv.org/abs/1606.01857}{{\tt 1606.01857}}].

\bibitem{Sarosi:2017ykf}
G.~Sárosi, \emph{{AdS$_{2}$ holography and the SYK model}},
  \href{http://dx.doi.org/10.22323/1.323.0001}{\emph{PoS} {\bf Modave2017}
  (2018) 001}, [\href{https://arxiv.org/abs/1711.08482}{{\tt 1711.08482}}].

\bibitem{Maldacena:2004sn}
J.~M. Maldacena, G.~W. Moore, N.~Seiberg and D.~Shih, \emph{{Exact vs.
  semiclassical target space of the minimal string}},
  \href{http://dx.doi.org/10.1088/1126-6708/2004/10/020}{\emph{JHEP} {\bf 10}
  (2004) 020}, [\href{https://arxiv.org/abs/hep-th/0408039}{{\tt
  hep-th/0408039}}].

\bibitem{Blommaert:2020seb}
A.~Blommaert, \emph{{Dissecting the ensemble in JT gravity}},
  \href{https://arxiv.org/abs/2006.13971}{{\tt 2006.13971}}.

\bibitem{McNamara:2019rup}
J.~McNamara and C.~Vafa, \emph{{Cobordism Classes and the Swampland}},
  \href{https://arxiv.org/abs/1909.10355}{{\tt 1909.10355}}.

\bibitem{Garcia-Etxebarria:2018ajm}
I.~García-Etxebarria and M.~Montero, \emph{{Dai-Freed anomalies in particle
  physics}}, \href{http://dx.doi.org/10.1007/JHEP08(2019)003}{\emph{JHEP} {\bf
  08} (2019) 003}, [\href{https://arxiv.org/abs/1808.00009}{{\tt 1808.00009}}].

\bibitem{GarciaEtxebarria:2020xsr}
I.~García~Etxebarria, M.~Montero, K.~Sousa and I.~Valenzuela, \emph{{Nothing
  is certain in string compactifications}},
  \href{https://arxiv.org/abs/2005.06494}{{\tt 2005.06494}}.

\bibitem{Milnor:1963}
J.~W. {Milnor}, \emph{{Spin structures on manifolds.}}, {\emph{{Enseign. Math.
  (2)}} {\bf 9} (1963) 198--203}.

\bibitem{Balasubramanian:2006sg}
V.~Balasubramanian, N.~Jokela, E.~Keski-Vakkuri and J.~Majumder, \emph{{A
  Thermodynamic interpretation of time for rolling tachyons}},
  \href{http://dx.doi.org/10.1103/PhysRevD.75.063515}{\emph{Phys. Rev. D} {\bf
  75} (2007) 063515}, [\href{https://arxiv.org/abs/hep-th/0612090}{{\tt
  hep-th/0612090}}].

\end{thebibliography}\endgroup

\end{document}